\pdfoutput=1
\documentclass[12pt,a4paper]{article}
\usepackage{ifthen} % for conditional statements
\newboolean{pdflatex}
\setboolean{pdflatex}{true} % False for eps figures 

\newboolean{articletitles}
\setboolean{articletitles}{true} % False removes titles in references

\newboolean{uprightparticles}
\setboolean{uprightparticles}{false} %True for upright particle symbols

\newboolean{inbibliography}
\setboolean{inbibliography}{false} %True once you enter the bibliography

\def\paperauthors{LHCb collaboration} % Leave as is for PAPER and CONF
\def\paperasciititle{Angular moments of the decay Lb to Lz  mu+ mu-} % Set ASCII title here
\def\papertitle{Angular moments of the decay \decay{\Lb}{\Lz\mumu} at low hadronic recoil} 
\def\paperkeywords{{High Energy Physics}, {LHCb}} % Comma separated
\def\papercopyright{\the\year\ CERN for the benefit of the LHCb collaboration} % new since 9/Apr/2018
\def\paperlicence{CC-BY-4.0 licence}
\def\paperlicenceurl{https://creativecommons.org/licenses/by/4.0/}

\usepackage[top=1in, bottom=1.25in, left=1in, right=1in]{geometry}

\usepackage{microtype}
\usepackage{lineno}  % for line numbering during review
\usepackage{xspace} % To avoid problems with missing or double spaces after
\usepackage{caption} %these three command get the figure and table captions automatically small

\usepackage{rotating}
\usepackage{graphicx}  % to include figures (can also use other packages)
\usepackage{color}
\usepackage{colortbl}
\graphicspath{{./figs/}} % Make Latex search fig subdir for figures

\usepackage{amsmath} % Adds a large collection of math symbols
\usepackage{amssymb}
\usepackage{amsfonts}
\usepackage{upgreek} % Adds in support for greek letters in roman typeset

\newcommand*\patchAmsMathEnvironmentForLineno[1]{%
\expandafter\let\csname old#1\expandafter\endcsname\csname #1\endcsname
\expandafter\let\csname oldend#1\expandafter\endcsname\csname
end#1\endcsname
 \renewenvironment{#1}%
   {\linenomath\csname old#1\endcsname}%
   {\csname oldend#1\endcsname\endlinenomath}%
}
\newcommand*\patchBothAmsMathEnvironmentsForLineno[1]{%
  \patchAmsMathEnvironmentForLineno{#1}%
  \patchAmsMathEnvironmentForLineno{#1*}%
}
\AtBeginDocument{%
\patchBothAmsMathEnvironmentsForLineno{equation}%
\patchBothAmsMathEnvironmentsForLineno{align}%
\patchBothAmsMathEnvironmentsForLineno{flalign}%
\patchBothAmsMathEnvironmentsForLineno{alignat}%
\patchBothAmsMathEnvironmentsForLineno{gather}%
\patchBothAmsMathEnvironmentsForLineno{multline}%
\patchBothAmsMathEnvironmentsForLineno{eqnarray}%
}

\usepackage{hyperxmp}

\usepackage[pdftex,
            pdfauthor={\paperauthors},
            pdftitle={\paperasciititle},
            pdfkeywords={\paperkeywords},
            pdfcopyright={Copyright (C) \papercopyright},
            pdflicenseurl={\paperlicenceurl}]{hyperref}

\usepackage[all]{hypcap} % Internal hyperlinks to floats.

\usepackage{xspace} 
\usepackage{upgreek}

\def\lhcb {\mbox{LHCb}\xspace}

\def\velo   {VELO\xspace}

\def\richone {RICH1\xspace}
\def\richtwo {RICH2\xspace}

\def\MagUp {\mbox{\em Mag\kern -0.05em Up}\xspace}

\ifthenelse{\boolean{uprightparticles}}%
{

 \def\Pmu         {\ensuremath{\upmu}\xspace}

 \def\Ppi         {\ensuremath{\uppi}\xspace}

 \def\Ppsi        {\ensuremath{\uppsi}\xspace}

 \def\PDelta      {\ensuremath{\Delta}\xspace}                 
 \def\PXi      {\ensuremath{\Xi}\xspace}                 
 \def\PLambda      {\ensuremath{\Lambda}\xspace}                 
 \def\PSigma      {\ensuremath{\Sigma}\xspace}                 
 \def\POmega      {\ensuremath{\Omega}\xspace}                 
 \def\PUpsilon      {\ensuremath{\Upsilon}\xspace}                 
 
 \def\PB      {\ensuremath{\mathrm{B}}\xspace}                 
                  
 \def\PD      {\ensuremath{\mathrm{D}}\xspace}

 \def\PJ      {\ensuremath{\mathrm{J}}\xspace}                 
 \def\PK      {\ensuremath{\mathrm{K}}\xspace}

 \def\Pb      {\ensuremath{\mathrm{b}}\xspace}                 
 \def\Pc      {\ensuremath{\mathrm{c}}\xspace}

 \def\Pi      {\ensuremath{\mathrm{i}}\xspace}

 \def\Pp      {\ensuremath{\mathrm{p}}\xspace}

 \def\Ps      {\ensuremath{\mathrm{s}}\xspace}

}
{

 \def\Pmu         {\ensuremath{\mu}\xspace}

 \def\Ppi         {\ensuremath{\pi}\xspace}

 \def\Ppsi        {\ensuremath{\psi}\xspace}                 
                  
 \mathchardef\PDelta="7101
 \mathchardef\PXi="7104
 \mathchardef\PLambda="7103
 \mathchardef\PSigma="7106
 \mathchardef\POmega="710A
 \mathchardef\PUpsilon="7107
                  
 \def\PB      {\ensuremath{B}\xspace}                 
                  
 \def\PD      {\ensuremath{D}\xspace}

 \def\PJ      {\ensuremath{J}\xspace}                 
 \def\PK      {\ensuremath{K}\xspace}

 \def\Pb      {\ensuremath{b}\xspace}                 
 \def\Pc      {\ensuremath{c}\xspace}

 \def\Pi      {\ensuremath{i}\xspace}

 \def\Pp      {\ensuremath{p}\xspace}

 \def\Ps      {\ensuremath{s}\xspace}

}

\makeatletter
\ifcase \@ptsize \relax% 10pt
  \newcommand{\miniscule}{\@setfontsize\miniscule{4}{5}}% \tiny: 5/6
\or% 11pt
  \newcommand{\miniscule}{\@setfontsize\miniscule{5}{6}}% \tiny: 6/7
\or% 12pt
  \newcommand{\miniscule}{\@setfontsize\miniscule{5}{6}}% \tiny: 6/7
\fi
\makeatother

\DeclareRobustCommand{\optbar}[1]{\shortstack{{\miniscule (\rule[.5ex]{1.25em}{.18mm})}
  \\ [-.7ex] $#1$}}

\def\mup        {{\ensuremath{\Pmu^+}}\xspace}
\def\mumu       {{\ensuremath{\Pmu^+\Pmu^-}}\xspace}

\def\squark    {{\ensuremath{\Ps}}\xspace}

\def\cquark    {{\ensuremath{\Pc}}\xspace}

\def\bquark    {{\ensuremath{\Pb}}\xspace}

\def\pion   {{\ensuremath{\Ppi}}\xspace}

\def\pip    {{\ensuremath{\pion^+}}\xspace}
\def\pim    {{\ensuremath{\pion^-}}\xspace}

\def\kaon    {{\ensuremath{\PK}}\xspace}
  \def\Kbar    {{\kern 0.2em\overline{\kern -0.2em \PK}{}}\xspace}

\def\KorKbar    {\kern 0.18em\optbar{\kern -0.18em K}{}\xspace}

\def\KS      {{\ensuremath{\kaon^0_{\mathrm{ \scriptscriptstyle S}}}}\xspace}

  \def\Dbar    {{\kern 0.2em\overline{\kern -0.2em \PD}{}}\xspace}

\def\DorDbar    {\kern 0.18em\optbar{\kern -0.18em D}{}\xspace}

\def\B       {{\ensuremath{\PB}}\xspace}
\def\Bbar    {{\ensuremath{\kern 0.18em\overline{\kern -0.18em \PB}{}}}\xspace}

\def\BorBbar    {\kern 0.18em\optbar{\kern -0.18em B}{}\xspace}
\def\Bz      {{\ensuremath{\B^0}}\xspace}

\def\jpsi     {{\ensuremath{{\PJ\mskip -3mu/\mskip -2mu\Ppsi\mskip 2mu}}}\xspace}

  \def\Y#1S{\ensuremath{\PUpsilon{(#1S)}}\xspace}% no space before {...}!

\def\proton      {{\ensuremath{\Pp}}\xspace}
\def\antiproton  {{\ensuremath{\overline \proton}}\xspace}

\def\Lz          {{\ensuremath{\PLambda}}\xspace}
\def\Lbar        {{\ensuremath{\kern 0.1em\overline{\kern -0.1em\PLambda}}}\xspace}
\def\LorLbar    {\kern 0.18em\optbar{\kern -0.18em \PLambda}{}\xspace}

\def\Lb      {{\ensuremath{\Lz^0_\bquark}}\xspace}
\def\Lbbar   {{\ensuremath{\Lbar{}^0_\bquark}}\xspace}

\newcommand{\decay}[2]{\ensuremath{#1\!\to #2}\xspace}         % {\Pa}{\Pb \Pc}

\def\to                 {\ensuremath{\rightarrow}\xspace}

\def\qsq       {{\ensuremath{q^2}}\xspace}

\def\CP                {{\ensuremath{C\!P}}\xspace}

\def\AT#1     {\ensuremath{A_{\mathrm{T}}^{#1}}\xspace}           % 2

\def\C#1      {\ensuremath{\mathcal{C}_{#1}}\xspace}                       % 9
\def\Cp#1     {\ensuremath{\mathcal{C}_{#1}^{'}}\xspace}                    % 7
\def\Ceff#1   {\ensuremath{\mathcal{C}_{#1}^{\mathrm{(eff)}}}\xspace}        % 9  
\def\Cpeff#1  {\ensuremath{\mathcal{C}_{#1}^{'\mathrm{(eff)}}}\xspace}       % 7
\def\Ope#1    {\ensuremath{\mathcal{O}_{#1}}\xspace}                       % 2
\def\Opep#1   {\ensuremath{\mathcal{O}_{#1}^{'}}\xspace}                    % 7

\newcommand{\tev}{\ifthenelse{\boolean{inbibliography}}{\ensuremath{~T\kern -0.05em eV}}{\ensuremath{\mathrm{\,Te\kern -0.1em V}}}\xspace}
\newcommand{\gev}{\ensuremath{\mathrm{\,Ge\kern -0.1em V}}\xspace}
\newcommand{\mev}{\ensuremath{\mathrm{\,Me\kern -0.1em V}}\xspace}
\newcommand{\kev}{\ensuremath{\mathrm{\,ke\kern -0.1em V}}\xspace}
\newcommand{\ev}{\ensuremath{\mathrm{\,e\kern -0.1em V}}\xspace}
\newcommand{\gevc}{\ensuremath{{\mathrm{\,Ge\kern -0.1em V\!/}c}}\xspace}
\newcommand{\mevc}{\ensuremath{{\mathrm{\,Me\kern -0.1em V\!/}c}}\xspace}
\newcommand{\gevcc}{\ensuremath{{\mathrm{\,Ge\kern -0.1em V\!/}c^2}}\xspace}
\newcommand{\gevgevcccc}{\ensuremath{{\mathrm{\,Ge\kern -0.1em V^2\!/}c^4}}\xspace}
\newcommand{\mevcc}{\ensuremath{{\mathrm{\,Me\kern -0.1em V\!/}c^2}}\xspace}

\def\mm   {\ensuremath{\mathrm{ \,mm}}\xspace}

\def\mum  {\ensuremath{{\,\upmu\mathrm{m}}}\xspace}

\def\invfb   {\ensuremath{\mbox{\,fb}^{-1}}\xspace}

\def\ps   {\ensuremath{{\mathrm{ \,ps}}}\xspace}

\newcommand{\stat}{\ensuremath{\mathrm{\,(stat)}}\xspace}
\newcommand{\syst}{\ensuremath{\mathrm{\,(syst)}}\xspace}

\newcommand{\chisq}{\ensuremath{\chi^2}\xspace}

\def\deriv {\ensuremath{\mathrm{d}}}

\def\gsim{{~\raise.15em\hbox{$>$}\kern-.85em
          \lower.35em\hbox{$\sim$}~}\xspace}
\def\lsim{{~\raise.15em\hbox{$<$}\kern-.85em
          \lower.35em\hbox{$\sim$}~}\xspace}

\def\sPlot{\mbox{\em sPlot}\xspace}

\def\sqs   {\ensuremath{\protect\sqrt{s}}\xspace}

\def\ptot       {\mbox{$p$}\xspace}
\def\pt         {\mbox{$p_{\mathrm{ T}}$}\xspace}

\def\mrad{\ensuremath{\mathrm{ \,mrad}}\xspace}

\def\evtgen     {\mbox{\textsc{EvtGen}}\xspace}

\def\geant      {\mbox{\textsc{Geant4}}\xspace}

\def\photos     {\mbox{\textsc{Photos}}\xspace}

\def\pythia     {\mbox{\textsc{Pythia}}\xspace}

\def\tell1  {TELL1\xspace}
\def\ukl1   {UKL1\xspace}

\newcommand{\eg}{\mbox{\itshape e.g.}\xspace}
\newcommand{\ie}{\mbox{\itshape i.e.}\xspace}

\def\AFBl {\ensuremath{A^{\ell}_{\mathrm{FB}}}\xspace}
\def\AFBh {\ensuremath{A^{h}_{\mathrm{FB}}}\xspace}
\def\AFBlh {\ensuremath{A^{\ell h}_{\mathrm{FB}}}\xspace}

\def\thetab {\ensuremath{\theta_b}\xspace}
\def\phib {\ensuremath{\phi_b}\xspace}
\def\thetal {\ensuremath{\theta_\ell}\xspace}
\def\phil {\ensuremath{\phi_\ell}\xspace}
\def\EOS {\texttt{EOS}\xspace}

\usepackage{cite} % Allows for ranges in citations
\usepackage{mciteplus}

\usepackage{longtable} % only for template; not usually to be used in PAPERs

\begin{document}

%%%%%%%%%%%%%%%%%%%%%%%%%
%%%%% Title     %%%%%%%%%
%%%%%%%%%%%%%%%%%%%%%%%%%
\renewcommand{\thefootnote}{\fnsymbol{footnote}}
\setcounter{footnote}{1}

% %%%%%%% CHOOSE TITLE PAGE--------
%\onecolumn

% $Id: title-LHCb-PAPER.tex 122526 2018-08-01 08:33:11Z kreps $
% ===============================================================================
% Purpose: LHCb-PAPER journal paper title page template
% Author: 
% Created on: 2010-09-25
% ===============================================================================

%%%%%%%%%%%%%%%%%%%%%%%%%
%%%%%  TITLE PAGE  %%%%%%
%%%%%%%%%%%%%%%%%%%%%%%%%
\begin{titlepage}
\pagenumbering{roman}

% Header ---------------------------------------------------
\vspace*{-1.5cm}
\centerline{\large EUROPEAN ORGANIZATION FOR NUCLEAR RESEARCH (CERN)}
\vspace*{1.5cm}
\noindent
\begin{tabular*}{\linewidth}{lc@{\extracolsep{\fill}}r@{\extracolsep{0pt}}}
\ifthenelse{\boolean{pdflatex}}% Logo format choice
{\vspace*{-1.5cm}\mbox{\!\!\!\includegraphics[width=.14\textwidth]{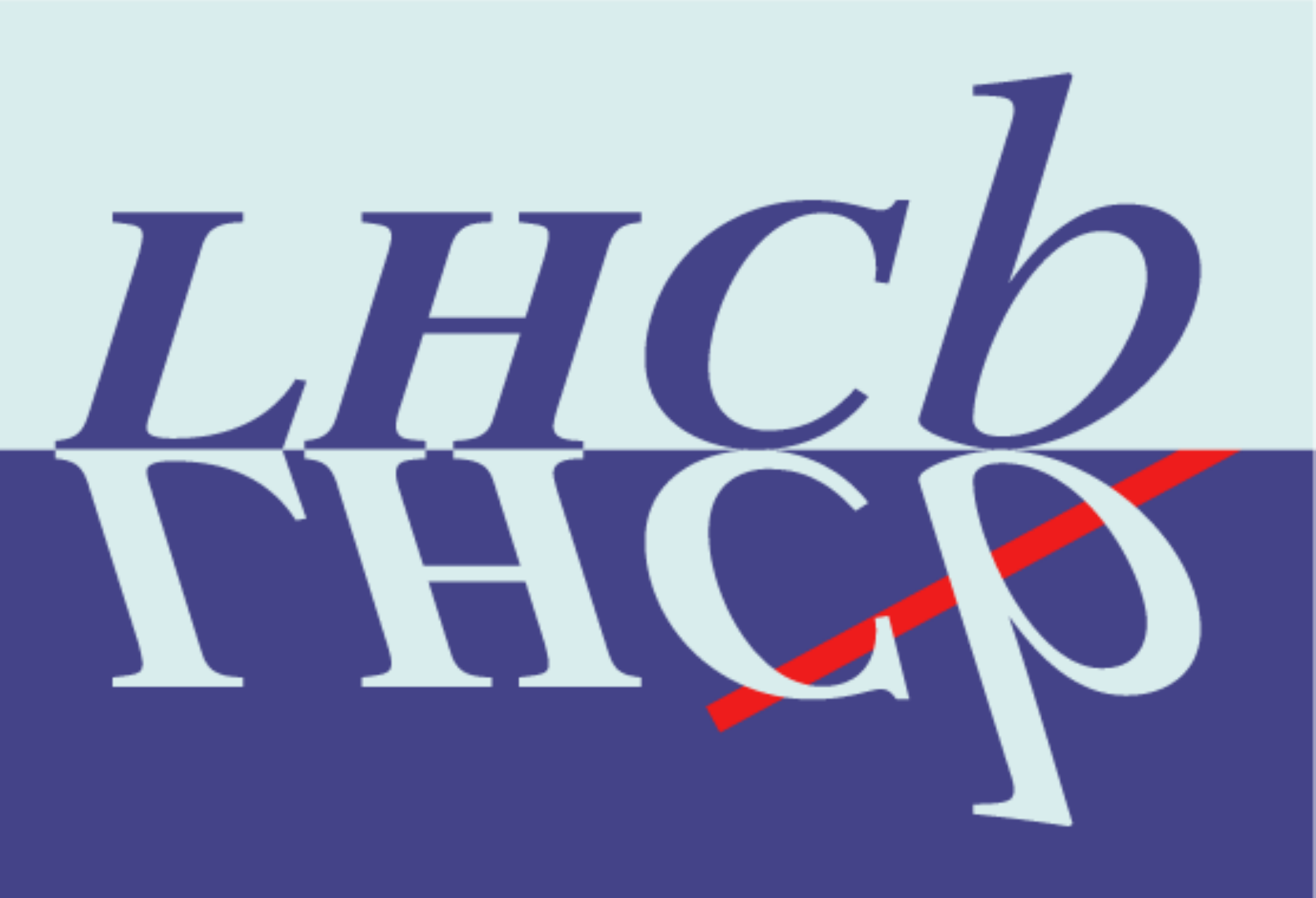}} & &}%
{\vspace*{-1.2cm}\mbox{\!\!\!\includegraphics[width=.12\textwidth]{lhcb-logo.eps}} & &}%
\\
 & & CERN-EP-2018-197 \\  % ID 
 & & LHCb-PAPER-2018-029 \\  % ID 
 & & August 1, 2018 \\ % Date - Can also hardwire e.g.: 23 March 2010
 & & \\
% not in paper \hline
\end{tabular*}

\vspace*{4.0cm}

% Title --------------------------------------------------
{\normalfont\bfseries\boldmath\huge
\begin{center}
% DO NOT EDIT HERE. Instead edit macro in main.tex to keep metadata correct
  \papertitle 
\end{center}
}

\vspace*{2.0cm}

% Authors -------------------------------------------------
\begin{center}
%In the footnote, replace 'paper' by 'Letter' in case of submission to PRL or PLB 
% Edit macro in main.tex to keep metadata correct
\paperauthors\footnote{Authors are listed at the end of this paper.}
\end{center}

\vspace{\fill}

% Abstract -----------------------------------------------
\begin{abstract}
  \noindent
An analysis of the angular distribution of the decay
\decay{\Lb}{\Lz\mumu} is presented, using data
collected with the LHCb detector between 2011 and 2016 and
corresponding to an integrated luminosity of approximately 5\invfb.
Angular observables are determined using a moment analysis of the
angular distribution at low hadronic recoil, 
corresponding to the dimuon invariant mass squared range $15 < \qsq < 20\gevgevcccc$.  
The full basis of observables is 
measured for the first time.  The lepton-side, hadron-side and
combined forward-backward asymmetries of the decay are determined to
be
\begin{displaymath}
\begin{split}
 \AFBl  & = -0.39  \pm 0.04\stat \,\pm 0.01 \syst~, \\ 
 \AFBh  & = -0.30  \pm 0.05\stat \,\pm 0.02 \syst~, \\ 
 \AFBlh & = +0.25  \pm 0.04\stat \,\pm 0.01 \syst~. 
\end{split}
\end{displaymath} 
The measurements are consistent with Standard Model predictions. 
\end{abstract}

\vspace*{2.0cm}

\begin{center}
  Published in JHEP 09 (2018) 146
\end{center}

\vspace{\fill}

{\footnotesize 
  % Edit macro in main.tex to keep metadata correct
\centerline{\copyright~\papercopyright. \href{\paperlicenceurl}{\paperlicence}.}}
\vspace*{2mm}

\end{titlepage}

%%%%%%%%%%%%%%%%%%%%%%%%%%%%%%%%
%%%%%  EOD OF TITLE PAGE  %%%%%%
%%%%%%%%%%%%%%%%%%%%%%%%%%%%%%%%

%  empty page follows the title page ----
\newpage
\setcounter{page}{2}
\mbox{~}
%\newpage
%

\cleardoublepage

%\twocolumn
% %%%%%%%%%%%%% ---------

\renewcommand{\thefootnote}{\arabic{footnote}}
\setcounter{footnote}{0}

%%%%%%%%%%%%%%%%%%%%%%%%%%%%%%%%
%%%%%  Table of Content   %%%%%%
%%%%%%%%%%%%%%%%%%%%%%%%%%%%%%%%
%%%% Uncomment next 2 lines if desired
%\tableofcontents
%\cleardoublepage

%%%%%%%%%%%%%%%%%%%%%%%%%
%%%%% Main text %%%%%%%%%
%%%%%%%%%%%%%%%%%%%%%%%%%

\pagestyle{plain} % restore page numbers for the main text
\setcounter{page}{1}
\pagenumbering{arabic}

%% Uncomment during review phase. 
%% Comment before a final submission.
%\linenumbers

% You can include short sections directly in the main tex file.
% However, for larger papers it is desirable to split the text into
% several semiautonomous files, which can be revised independently.
% This is especially useful when developing a document in
% collaboration with several people, since then different parts can be
% edited independently.  This type of file organization is shown here.
% 

\section{Introduction}
\label{sec:Introduction}

In the Standard Model of particle physics (SM), the decay
\decay{\Lb}{\Lz\mumu} proceeds via a \bquark to \squark quark
flavour-changing neutral-current transition.  The decay is
consequently rare in the SM, with a branching fraction of order
$10^{-6}$~\cite{Detmold:2016pkz}.  In extensions of the SM the branching
fraction and angular distribution 
of the decay can be modified significantly, with the latter providing a large number
 of particularly sensitive observables (see \eg Ref.\ \cite{Boer:2014kda}).
 The rate and angular distribution of corresponding \B meson decays have been studied by the \B-factory experiments, CDF at the TeVatron and the ATLAS, CMS and LHCb experiments at the LHC. 
 A global analyses of the measurements favours a modification of the coupling strengths of the \bquark to \squark transition from their SM values at the level of 4 to 5 standard deviations~\cite{Altmannshofer:2017fio,Ciuchini:2017mik,Chobanova:2017ghn,Geng:2017svp,Capdevila:2017bsm}.
 The decay \decay{\Lb}{\Lz\mumu} has several important phenomenological differences to the \B meson decays: 
 the \Lb baryon is a spin-half particle and could be produced polarised; 
 the transition involves a diquark system as a spectator, rather than
 a single quark;
 and the \Lz baryon decays weakly resulting in observables related to the hadronic part of the decay that are not present in the meson decays. 
 The decay \decay{\Lb}{\Lz\mumu} therefore provides an important additional test of the SM predictions, which can be used to improve our understanding of the nature of the anomalies seen in the \B meson decays. 

The decay \decay{\Lb}{\Lz\mumu} was first observed by the CDF
collaboration \cite{Aaltonen:2011qs}.  The LHCb collaboration has subsequently studied 
the rate of the decay  as a function of the dimuon invariant mass squared,
\qsq, in Refs.~\cite{LHCb-PAPER-2013-025,LHCb-PAPER-2015-009}.  In the LHCb
analysis, evidence for a signal was only found at low hadronic recoil
(corresponding to the range $15 < \qsq < 20\gev^{2}/c^{4}$).  
This is consistent with recent SM predictions based on Lattice QCD calculations of the form factors of the decay~\cite{Detmold:2016pkz}.
The angular distribution of the decay was studied for the first time
in Ref.~\cite{LHCb-PAPER-2015-009}, using two projections of the
five-dimensional angular distribution of the decay and a data set corresponding to an integrated luminosity of 3\invfb.  
The analysis measured two angular asymmetries using the hadronic and
leptonic parts of the decay in the range $15 < \qsq < 20\gev^2/c^4$.

This paper presents the first measurement of the full basis of angular
observables for the decay \decay{\Lb}{\Lz\mumu} in the range $15 < \qsq < 20\gev^{2}/c^4$.%
\footnote{The
  inclusion of charge-conjugated processes is implicit throughout.
  }
The measurement uses
$pp$ collision data, corresponding to an integrated luminosity of
approximately 5\invfb, collected between 2011 and 2016 at
centre-of-mass energies of 7, 8 and 13\tev.
The paper is organised as follows: Sec.~\ref{sec:Moments} introduces
the moment analysis used to characterise the angular observables;
Sec.~\ref{sec:Detector} describes the LHCb detector;
Sec.~\ref{sec:Selection} outlines the selection of
\decay{\Lb}{\Lz\mumu} candidates, where the \Lz is reconstructed in the \proton\pim final state;
Sec.~\ref{sec:Yields} presents the fit to the invariant-mass
distribution of $p\pim\mumu$ candidates, from which the yield of
the \decay{\Lb}{\Lz\mumu} signal is obtained; results
are given in Sec.~\ref{sec:Results};  Section~\ref{sec:Systematics}
summarises potential sources of systematic uncertainty; and conclusions are presented in
Sec.~\ref{sec:Summary}.

\section{Moments of the angular distribution}
\label{sec:Moments}

The angular distribution of the \decay{\Lb}{\Lz\mumu} decay can be
described using a normal unit-vector, $\hat{n}$, defined by the
vector product of the beam direction and the \Lb momentum vector, and five
angles \cite{Blake:2017une}: 
the angle, $\theta$, between $\hat{n}$ and the \Lz 
baryon direction in the rest frame of the \Lb baryon;
polar and azimuthal angles \thetal\ and \phil\ describing the decay of the
dimuon system; and polar and azimuthal
angles \thetab\ and \phib\ describing the decay of the \Lz baryon.
An explicit definition of the angular basis is provided in Appendix~\ref{appendix:Basis}. 
The beam direction is assumed to be aligned with the positive $z$ direction in the LHCb coordinate system~\cite{Alves:2008zz}.\footnote{The coordinate system is defined with the centre of the LHCb vertex detector as the origin and positive $z$ pointing along the beam-line in the direction of the detector's dipole magnet.}  
The small crossing angle of the colliding beams is neglected in the analysis but is considered as a source of systematic uncertainty.
If the \Lb baryon is produced without any preferred
polarisation, the angular distribution only depends on the angles $\theta_\ell$ and $\theta_b$ and on the angle between the decay planes of the \Lz baryon and the
dimuon system ($\phil + \phib$).  An illustration of this angular basis
can be found in Ref.~\cite{Blake:2017une}.

The full angular distribution, averaged over the range $15 < \qsq < 20\gev^2/c^4$, can be described by the sum of 34 \qsq-dependent angular
terms~\cite{Blake:2017une},
\begin{align}
  \frac{\deriv^5\Gamma}{\deriv\vec{\Omega}} = \frac{3}{32\pi^{2}}
  \sum\limits_{i}^{34} K_{i} f_{i}(\vec{\Omega})~,
  \label{eq:terms}
\end{align}
where $\vec{\Omega}\equiv(\cos\theta,\cos\thetal,\phil,\cos\thetab,\phib)$ and the $f_{i}(\vec{\Omega})$ functions have different dependencies on the angles.
The $K_{i}$ parameters depend on the underlying short-distance physics and on the
form factors governing the $\Lb\to\Lz$ transition. 
The full form of the distribution is given in Appendix~\ref{appendix:distribution}.
Equation~\ref{eq:terms} is normalised such that $2 K_1 + K_2 = 1$.
Twenty-four of the
observables, $K_{11}$ to $K_{34}$, are proportional to the \Lb production polarisation and are zero if the \Lb baryons are produced
unpolarised. The reduced form of the angular distribution in the
case of zero production polarisation can be found in
Refs.~\cite{Gutsche:2013pp,Boer:2014kda}.

The $K_{i}$ parameters can be determined from data by means of a maximum-likelihood fit or via a moment analysis~\cite{James:2006zz,Beaujean:2015xea}.  The latter is
preferred in this analysis due to the small size of the available data
sample and the large number of unknown parameters.  
To determine the values of the $K_{i}$ parameters, weighting functions
$g_i(\vec{\Omega})$ are chosen to project out individual angular observables.
The $g_i(\vec{\Omega})$ functions, which are orthogonal to the
$f_j(\vec{\Omega})$ functions, 
are normalised such that
\begin{align}
K_{i} &= \int \frac{\deriv^5\Gamma}{\deriv\vec{\Omega}}
g_{i}(\vec{\Omega}) \deriv\vec{\Omega}~.
\end{align} 
The set of weighting functions used in this analysis can be found in Refs.~\cite{Blake:2017une,Beaujean:2015xea} and listed in Appendix~\ref{appendix:distribution}.
For the case of ideal detector response and in the absence
of background, the $K_{i}$ parameters can be estimated from data by summing
over the observed candidates.  In realistic scenarios, per-candidate weights are
necessary to compensate for nonuniform selection efficiency
and background contamination. 
The $K_{i}$ parameters are then estimated as
\begin{align}
 K_{i} = \sum\limits_{n} w_n \, g_{i}(\vec{\Omega}_{n})
\Big/ \sum\limits_{n} w_n~,
\end{align} 
where $w_n$ is the product of the two weights associated with candidate $n$.  The
background is subtracted using weights based on the \emph{sPlot}
technique~\cite{Pivk:2004ty,Xie:2009rka}.  The efficiency to
reconstruct and select the candidates is determined using samples of
simulated events.  The small effects of finite angular resolution are neglected in the
analysis but are considered as a source of systematic uncertainty.

\section{Detector and simulation}
\label{sec:Detector}

The \lhcb detector~\cite{Alves:2008zz,LHCb-DP-2014-002} is a
single-arm forward spectrometer covering the \mbox{pseudorapidity}
range $2<\eta <5$, designed for the study of particles containing
\bquark or \cquark quarks. The detector includes a high-precision
tracking system consisting of a silicon-strip vertex detector (\velo)
surrounding the $pp$ interaction region~\cite{LHCb-DP-2014-001}, a
large-area silicon-strip detector located upstream of a dipole magnet
with a bending power of about $4{\mathrm{\,Tm}}$, and three stations
of silicon-strip detectors and straw drift
tubes~\cite{LHCb-DP-2013-003} placed downstream of the magnet.  The
tracking system provides a measurement of the momentum, \ptot, of charged
particles with a relative uncertainty that varies from 0.5\% at low
momentum to 1.0\% at 200\gevc.  The minimum distance between a track
and a primary $pp$ interaction vertex (PV), the impact parameter (IP),
is measured with a resolution of $(15+29/\pt)\mum$, where \pt is the
component of the momentum transverse to the beam, in\,\gevc.
Different types of charged hadrons are distinguished using information
from two ring-imaging Cherenkov detectors (\richone and \richtwo)
\cite{LHCb-DP-2012-003}.
Photons, electrons and hadrons are identified by a calorimeter system
consisting of scintillating-pad and preshower detectors, an
electromagnetic calorimeter and a hadronic calorimeter. Muons are
identified with a system composed of alternating layers of iron and
multiwire proportional chambers~\cite{LHCb-DP-2012-002}.

The online event selection is performed by a
trigger~\cite{LHCb-DP-2012-004}, which consists of a hardware stage,
based on information from the calorimeter and muon systems, followed
by a software stage, which applies a full event reconstruction.  The
signal candidates are required to pass through a hardware trigger that
selects events containing at least one muon with large $\pt$ or a pair of muons with a large product of their transverse momenta.
The \pt threshold of the single muon trigger varied in the range between 1 and 2\gevc, depending on the
data-taking conditions.  The subsequent software trigger requires a two-, three-
or four-track secondary vertex with a significant displacement from
any PV.  At least one of the tracks 
must have a transverse momentum $\pt > 1\gevc$ and be
inconsistent with originating from a PV.  A multivariate
algorithm~\cite{BBDT} is used to identify whether the secondary
vertex is consistent with the decay of a \bquark hadron.
 
Samples of simulated \decay{\Lb}{\Lz\mumu} events are used to develop
an offline event selection and to quantify the effects of detector
response, candidate reconstruction and selection on the measured
angular distribution.  In the simulation, $pp$ collisions are
generated using \pythia~\cite{Sjostrand:2007gs,Sjostrand:2006za} with
a specific \lhcb configuration~\cite{LHCb-PROC-2010-056}.  Decays of
hadrons are described by \evtgen~\cite{Lange:2001uf}, in
which final-state radiation is generated using
\photos~\cite{Golonka:2005pn}.  The interaction of the generated
particles with the detector, and its response, are implemented using
the \geant toolkit~\cite{Allison:2006ve, *Agostinelli:2002hh} as
described in Ref.~\cite{LHCb-PROC-2011-006}.  The samples of simulated
data are corrected to account for observed differences relative to
data in detector occupancy, vertex quality and the production kinematics of the \Lb baryon.
The particle identification performance of the detector is measured
using calibration samples of data.

\section{Candidate selection}
\label{sec:Selection}

Signal candidates are formed by combining a \Lz baryon candidate with
two oppositely charged particles that are identified as muons by the
muon system and have track segments in the \velo. 
Only muon pairs with \qsq in the range $15 < \qsq <
20\gevgevcccc$, where the majority of the \decay{\Lb}{\Lz\mumu} signal
is expected to be observed, are considered.  Candidates in the range
$8 < \qsq < 11\gevgevcccc$, which predominantly consist of decays via
an intermediate \jpsi meson that subsequently decays to \mumu, are
also retained and used to cross-check various aspects of the
analysis.

 Candidate \Lz decays are reconstructed in the
 \decay{\Lz}{\proton\pim} decay mode from two oppositely charged tracks.
The tracks are reconstructed in one of two categories, depending on where the \Lz decayed in the detector. 
The two tracks either both include information from the \velo (\textit{long} candidates) or both do not include
 information from the \velo (\textit{downstream} candidates).\footnote{Tracks with
   information from the \velo typically have a better momentum
   resolution and are associated with \Lz baryons with shorter
   lifetimes.} 
 The \Lz candidates must also have: a vertex fit with a good \chisq; 
 a decay time of at least 2\ps; an invariant mass within 30\mevcc of the known \Lz
 mass~\cite{PDG2018}; and a decay vertex at $z < 2350\mm$.
The requirement on the decay position removes background from hadronic
interactions in the material at the exit of the \richone detector.
The \Lz baryon and the dimuon pair are required to form a 
vertex with a good fit quality.  The resulting \Lb candidate
is required to be consistent with originating from one of the PVs in the
event and to have a vertex position that is significantly displaced from
that PV.

An artificial neural network is trained to further suppress combinatorial
background, in which tracks from an event are mistakenly combined to
form a candidate.  The neural network uses simulated
\decay{\Lb}{\Lz\mumu} decays as a proxy for the signal and candidates
from the upper mass sideband of the data, with a $\Lz\mumu$ invariant
mass greater than 5670\mevcc, for the background.  The inputs to the
neural network are: the \chisq of the vertex fit to the \Lb candidate;
the \Lb decay-time and the angle between the \Lb momentum
vector and the vector between the PV and the \Lb decay vertex; the \Lz
flight distance from the PV, its \pt and reconstructed mass; the IP of
the muon with the highest \pt; the IP of either the pion or proton
from the \Lz, depending on which has the highest \pt; and a measure of
the isolation of the \Lb baryon in the detector.  The working point of
the neural network is chosen to maximise the expected significance of the
\decay{\Lb}{\Lz\mumu} signal in the $15 < \qsq < 20\gevgevcccc$
region, assuming the branching fraction measured in Ref.~\cite{LHCb-PAPER-2015-009}. 
It is checked that selecting events based on their neural network response does not introduce any significant bias in the reconstructed $p\pim\mumu$ mass distribution, $m(p\pim\mumu)$.

\section{Candidate yields}
\label{sec:Yields}

\begin{figure}[tb]
\centering
\includegraphics[width=0.49\linewidth]{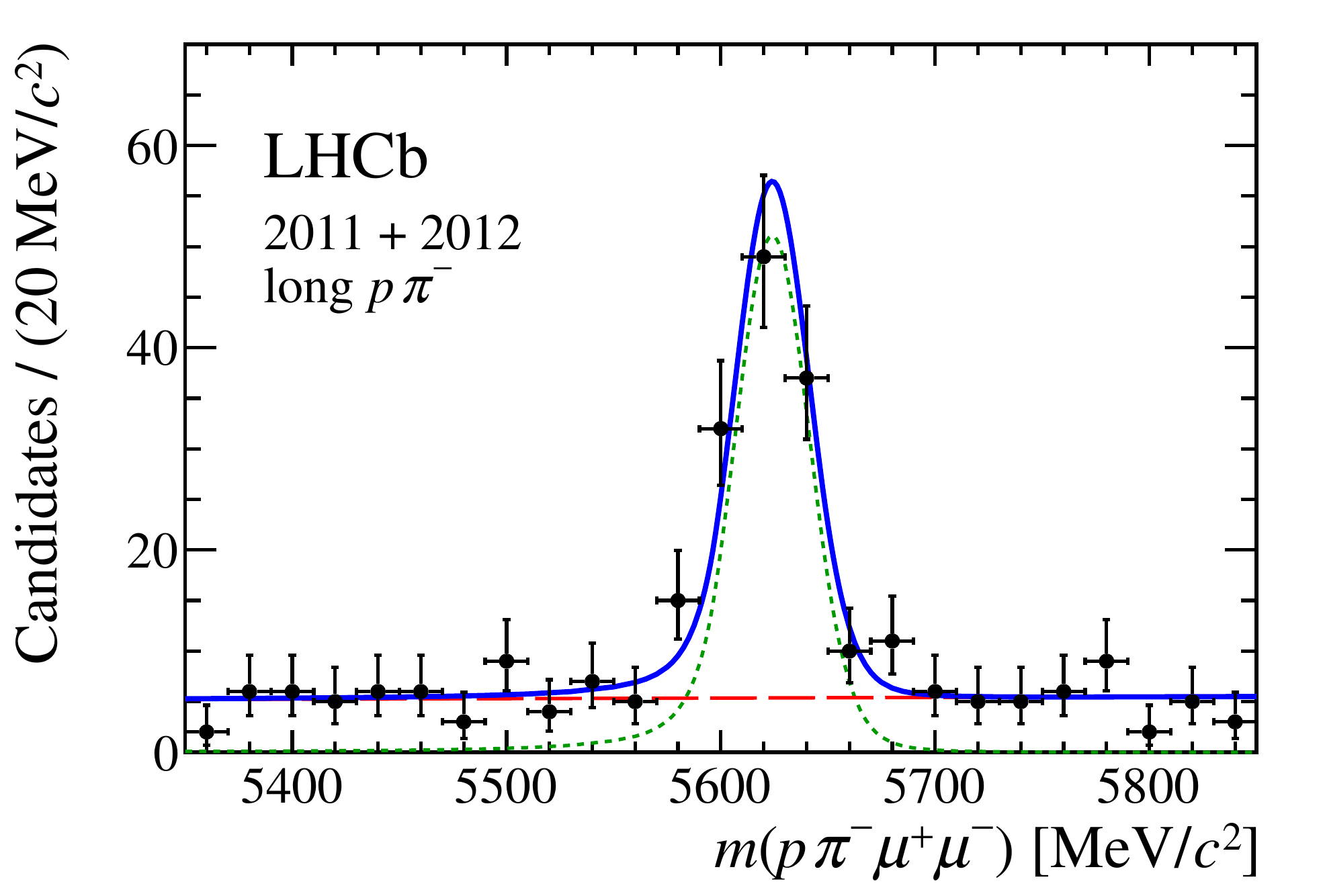} 
\includegraphics[width=0.49\linewidth]{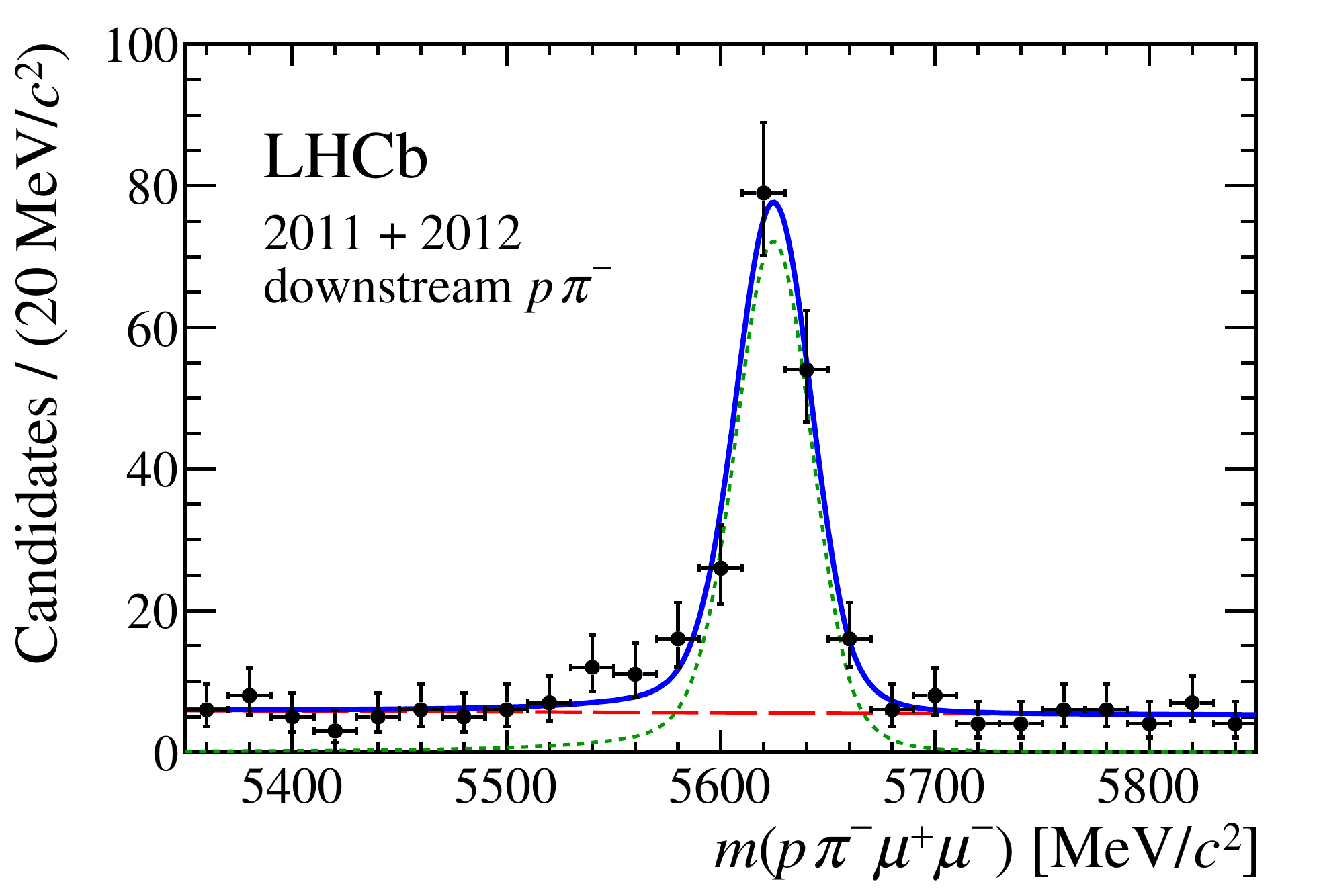}  \\ 
\includegraphics[width=0.49\linewidth]{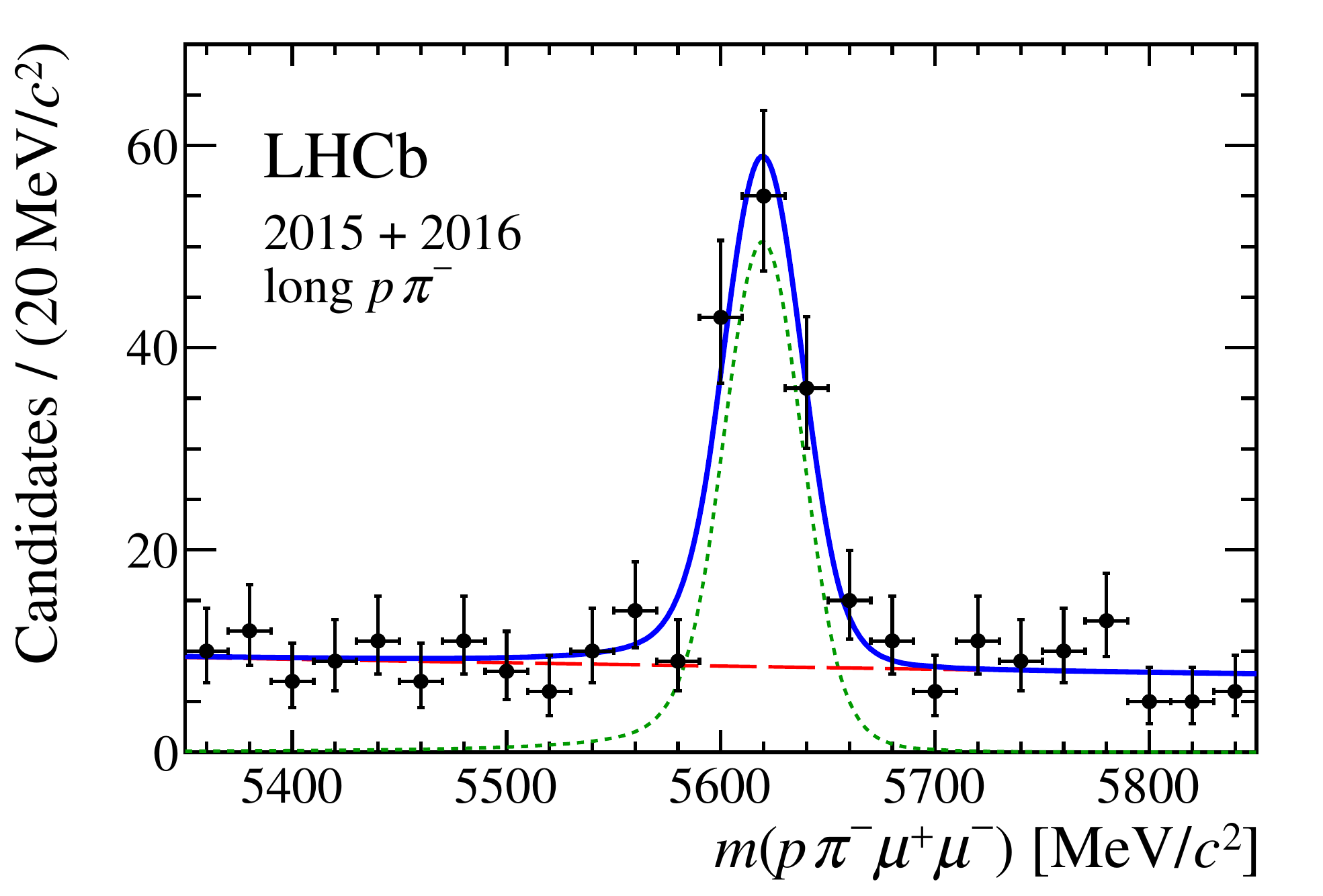} 
\includegraphics[width=0.49\linewidth]{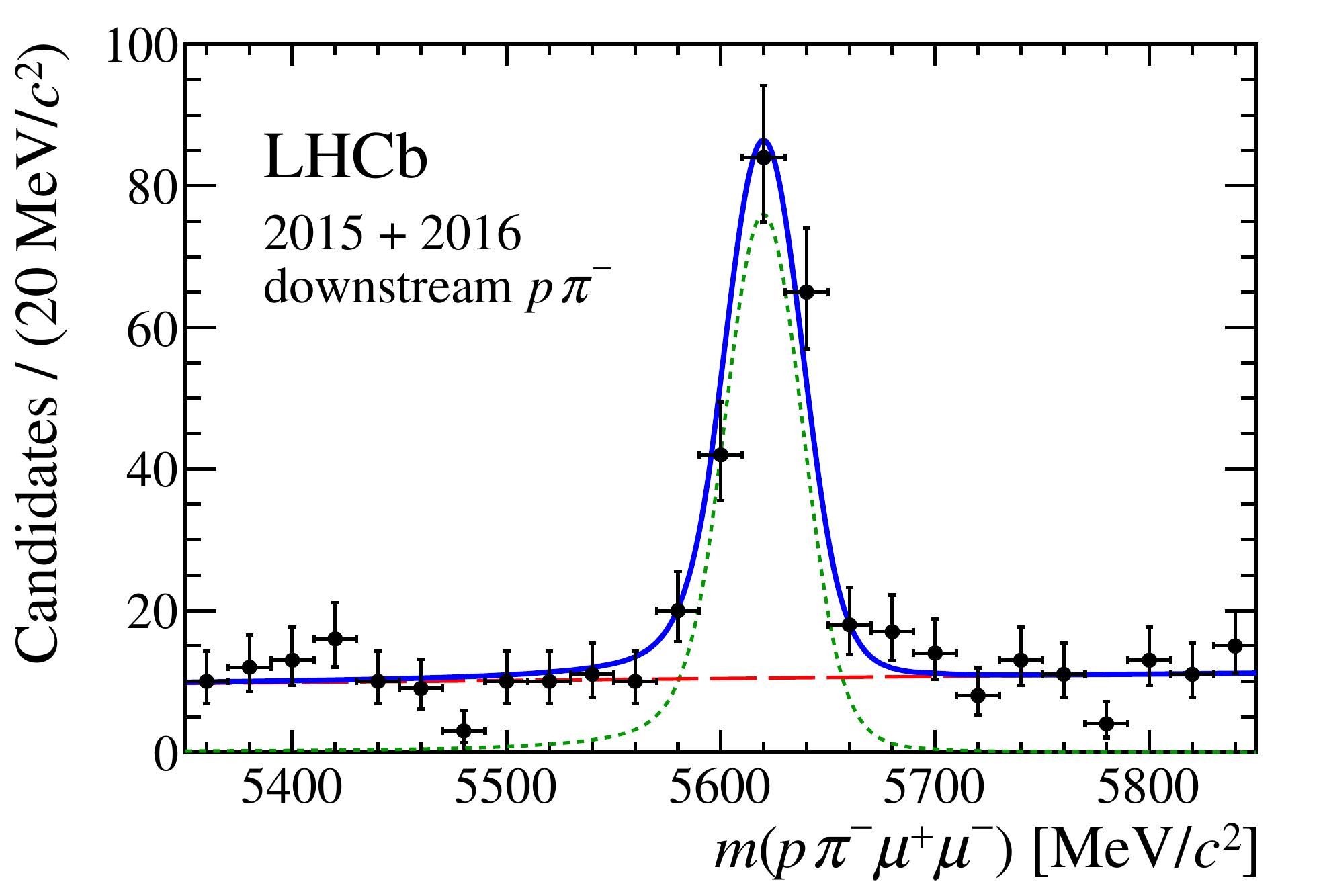}  
\caption{ Distribution of $p\pim\mumu$ invariant mass for (left)
  long- and (right) downstream-track $p\pim$ categories in the (top) 
  Run\,1 data and (bottom) Run\,2 data.  The result of the
  fit to each sample of data is indicated by the solid blue line.  The signal and
  background components are illustrated by the dotted green and dashed red
  lines, respectively.}
\label{fig:yields}
\end{figure}

Figure~\ref{fig:yields} shows the $p\pim\mumu$ mass distribution of the selected candidates in the Run\,1 and Run\,2 data sets, separated into the long-track and downstream-track $p\pim$ categories. 
The candidates comprise a mixture of \decay{\Lb}{\Lz\mumu} decays, combinatorial background and a negligible
contribution from other \bquark-hadron decays.  The
largest single component of the latter arises from the decay
\decay{\Bz}{\KS\mumu}, where the \KS meson decays to $\pip\pim$ and is mis-reconstructed 
as a \Lz baryon.  

The yield of \decay{\Lb}{\Lz\mumu} decays is determined by
performing an unbinned extended maximum-likelihood fit to $m(p\pim\mumu)$.  
In the fit, the signal is described by the sum of two modified Gaussian functions, one with a power-law tail on the low-mass side and the other with a power-law tail on the high-mass side of the distribution. 
The two Gaussian functions have a common peak position and width parameter but
different tail parameters and relative fractions.
The tail parameters and the relative fraction of the two 
functions is fixed from fits performed to simulated
\decay{\Lb}{\Lz\mumu} decays.  The mean and width are determined from
fits to \decay{\Lb}{\jpsi\Lz} candidates in the data.  
A small correction is applied to the width parameter to account for a
\qsq dependence of the resolution seen in the simulation. 
Combinatorial
background is described by an exponential function, with a slope
parameter that is determined from data.  The parameters
describing the signal and the background are determined separately for
each data-taking period and for the long- and the downstream-track $p\pim$
categories.

The fits result in yields of $120\pm 13$ ($175\pm 15$) and $126\pm 13$ ($189\pm 16$ ) decays in the long (downstream) $p\pim$ category of the Run\,1 and Run\,2 data, respectively. 
These fits are  used to the determine the weights needed to subtract the background in the moment analysis.
The yields are consistent with those expected based on the estimated signal efficiency, the recorded integrated luminosity and the scaling of the \Lb production cross-section with centre-of-mass energy.

\section{Angular efficiency}
\label{sec:Efficiency}

The trigger, reconstruction and the selection process distort
the measured angular distribution of the \decay{\Lb}{\Lz\mumu} decays.
The largest distortions are found to be the result of kinematic
requirements in the reconstruction, most notably due to an implicit
momentum threshold applied by requiring that the muons traverse the detector and
reach the muon system.  The angular efficiency is parameterised in six
dimensions taking into account the correlations between the different
angles and the \qsq-dependence of the angular efficiency.
\begin{align}
\varepsilon(\vec{\Omega},\qsq) = \sum_{ijmnrs} c_{ijmnrs}\,
L_i(\cos\theta)\,L_j(\cos\theta_\ell)\, L_m(\phi_\ell/\pi)\,
L_n(\cos\theta_b) \,L_r(\phi_b/\pi)\, L_s(\qsq)~,
\label{eq:efficiency}
\end{align} 
where the $L_{t}(x)$ denote a Legendre polynomial of order $t$ in
variable $x$, and the \qsq range considered has been rescaled linearly
between $-1$ and $+1$.  The
coefficients $c_{ijmnrs}$ are determined by performing a moment
analysis of \decay{\Lb}{\Lz\mumu} decays simulated according to a phase-space model. 
The simulated decays are weighted such that they are uniformly distributed in \qsq and in the five angles, 
after which the angular distribution of the selected decays is proportional to the efficiency. 

To achieve a good parameterisation of the efficiency, a large number of terms is required. 
The number of terms is reduced using an iterative approach. 
As a first step, the efficiency projection of each variable is
parameterised independently using the sum of Legendre polynomials of
up to eighth order.
As a second step, correlations between pairs of angles and 
between individual angles and \qsq are accounted for in turn. 
These corrections are parameterised by sums involving pairs of polynomials that run up to sixth order in each variable. 
As a final step, a six-dimensional correction is
applied allowing for  polynomials of up to
first order in the angles and \qsq. 
Before each step, the simulated decays are corrected to remove the effects parameterised in the previous step. 
Small differences in the efficiency to reconstruct \proton/\antiproton and \pip/\pim are neglected. 

The angular efficiency model is cross-checked in data using
\mbox{\decay{\Lb}{\jpsi\Lz}} and \mbox{\decay{\Bz}{\jpsi\KS}} decays, with
\mbox{\decay{\jpsi}{\mumu}}.
These decays have a similar topology to the \mbox{\decay{\Lb}{\Lz\mumu}} decay and well known angular distributions. 
For the \mbox{\decay{\Bz}{\jpsi\KS}} decay, where the \KS decays to \pip\pim, the parameter $K_1$ is one-half and the remaining observables are equal to zero. 
The angular distribution of the \mbox{\decay{\Lb}{\jpsi\Lz}} decay is compatible with the measurements in Refs.~\cite{LHCb-PAPER-2012-057,Aad:2014iba,Sirunyan:2018bfd}.

\section{Results}
\label{sec:Results}

The angular observables are obtained using a moment analysis of the angular distribution, weighting candidates as described in Sec.~\ref{sec:Efficiency} to account for their detection efficiency.
Background is subtracted using weights obtained from the \sPlot
technique from the fits described in Sec.~\ref{sec:Yields}.\ The weights used to correct for the efficiency and subtract the background are determined separately for each data-taking period and for the long-track and downstream-track $p\pim$ categories. 
The $K_i$ parameters are then determined from a data set that combines the two reconstruction categories. 
As the polarisation of the \Lb baryons at production may vary with 
 centre-of-mass energy between the Run\,1 data, collected at
$\sqs=7$ and 8\tev, and the Run\,2 data, collected at
$\sqs=13\tev$, these two data sets are initially treated independently.  
The results for the two data-taking periods are given in Appendix~\ref{appendix:results:split}.  
The statistical uncertainties on the various $K_{i}$ parameters are determined
using a bootstrapping technique~\cite{Efron:1979}.  
In each step of the bootstrap, the process of subtracting the background and the weighting of the candidates is repeated.

A \chisq comparison of the results from the two data-taking periods, taking into account the correlations between the
 observables, yields a \chisq of 35.0 with 33 degrees of
freedom.  This indicates an excellent agreement between the two data
sets and suggests that the production polarisation is consistent for
the centre-of-mass energies studied.  The Run\,1 and Run\,2 data samples 
are therefore combined and the observables are determined on the combined sample.  The results are given in Table~\ref{tab:results}.
The correlation between the angular observables is presented in Appendix~\ref{appendix:correlation}.
Figure~\ref{fig:projections} shows the one-dimensional angular
projections of $\cos\thetal$, $\cos\thetab$, $\cos\theta$, \phil
and \phib for the background-subtracted candidates.  The
data are described well by the product of the angular distributions
obtained from the moment analysis and the efficiency model.

\begin{figure}[!tb]
\centering
\includegraphics[width=\linewidth]{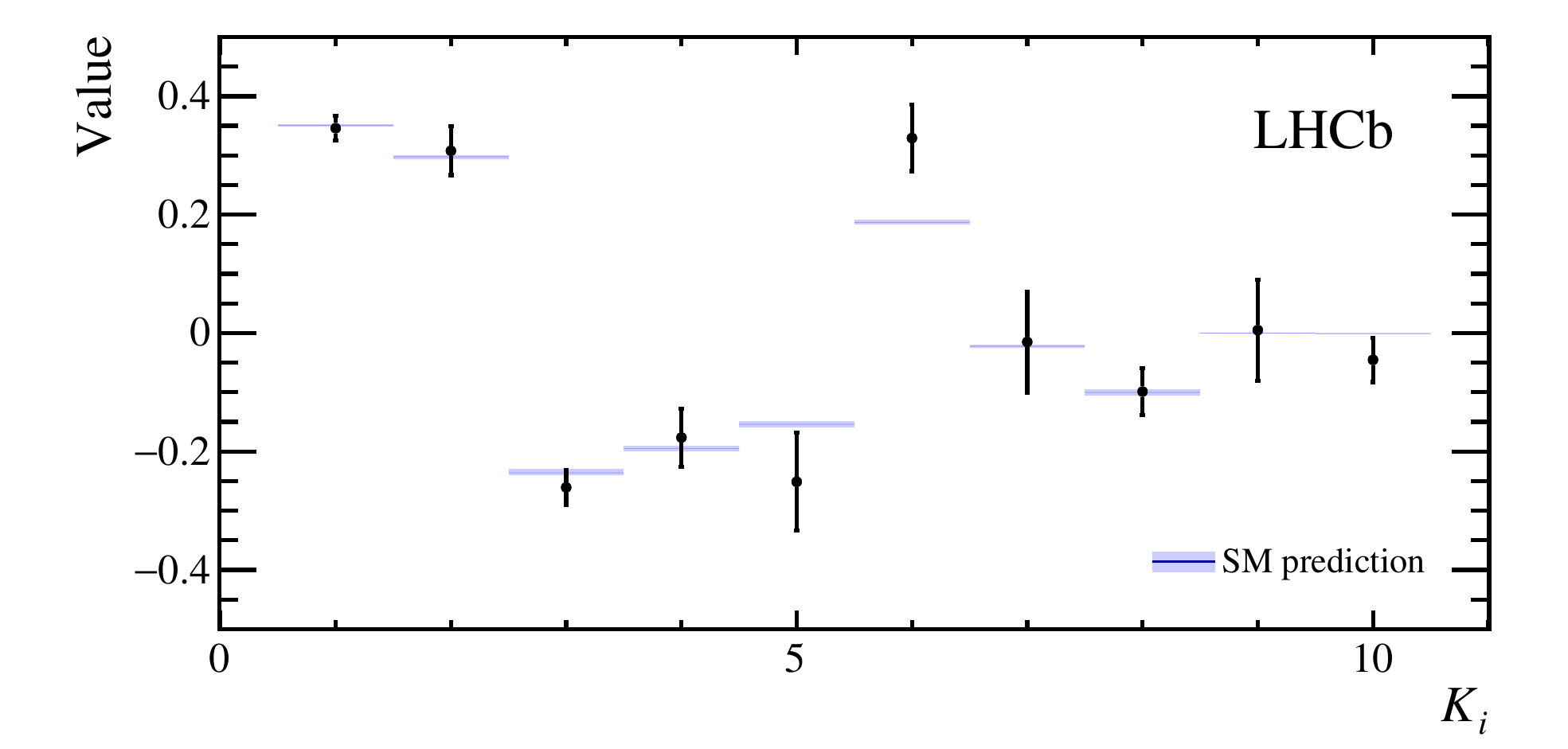} 
\caption{Angular observables combining the results for the moments obtained
from Run\,1 and Run\,2 data, as well as candidates reconstructed in
the long- and downstream-track $\proton\pim$ categories.
The blue line represents the SM predictions obtained using the \EOS software. The thickness of the light-blue band represents the uncertainty on the SM predictions. 
}
\label{fig:results}
\end{figure}

Figure~\ref{fig:results} compares the measured observables with their
corresponding SM predictions, obtained from the \EOS software~\cite{OurEOSVersion} using the values of the \Lb production
polarisation measured in Ref.\ \cite{LHCb-PAPER-2012-057}.
The values of the observables $K_{11}$ to $K_{34}$ are
consistent with zero.  This is expected from measurements of the
angular distribution of the decay \decay{\Lb}{\jpsi\Lz} by CMS~\cite{Sirunyan:2018bfd} and
LHCb~\cite{LHCb-PAPER-2012-057}, which indicate that the
production polarisation of \Lb baryons is small in $pp$ collisions at
7 and 8\tev.
The measurements are consistent with the SM predictions for $K_{1}$ to $K_{10}$. 
The largest discrepancy is seen in $K_6$, which is 2.6 standard deviations from the SM prediction. 
The angular observables result in an angular distribution that is not positive for all values of the angles. 
To obtain a physical angular distribution, $K_6$ has to move closer to its SM value. 
%There is also a consistency between the measurements and the $K_i$ values predicted in the new physics scenarios favoured by the global fits to the \bquark to \squark quark data~\cite{Altmannshofer:2017fio,Ciuchini:2017mik,Chobanova:2017ghn,Geng:2017svp,Capdevila:2017bsm}.
The measured $K_i$ values are also consistent with the values predicted by new physics scenarios favoured by global fits to data from \bquark to \squark quark transitions~\cite{Altmannshofer:2017fio,Ciuchini:2017mik,Chobanova:2017ghn,Geng:2017svp,Capdevila:2017bsm}.
These new physics scenarios result in only a small change of $K_{1}$ to $K_{10}$ in the low-recoil region. 

The $K_{i}$ observables can be combined to determine the angular asymmetries
\begin{displaymath}
\begin{split}
\AFBl = \tfrac{3}{2} K_{3} 
        & = -0.39 \pm 0.04 \,\pm 0.01~, \\ 
\AFBh = K_{4} +  \tfrac{1}{2} K_{5} 
        & = -0.30 \pm 0.05 \,\pm 0.02~, \\ 
\AFBlh = \tfrac{3}{4} K_{6}
        &=  +0.25 \pm 0.04 \,\pm 0.01~, 
\end{split}
\end{displaymath}
where the first uncertainties are statistical and the second are the systematic
uncertainties that are discussed in the following section.  The
forward-backward asymmetries \AFBl and \AFBh are in good agreement
with the SM predictions.  The asymmetry \AFBlh, which is proportional
to $K_6$, is 2.6 standard deviations from its SM prediction.
The value of \AFBh is consistent with that measured in Ref.~\cite{LHCb-PAPER-2015-009}. 
The value of \AFBl is not comparable due to an inconsistency in the definition of $\theta_{\ell}$ in that reference.\footnote{
Under the definition of $\theta_{\ell}$ used in Ref.~\cite{LHCb-PAPER-2015-009},  \AFBl measured the asymmetry difference between \Lb and \Lbbar decays rather than the average of the asymmetries.
}

\begin{table}[!tb]
\caption{
Angular observables combining the results of the moments obtained
from  Run\,1 and Run\,2 data, as well as candidates reconstructed in
the long- and downstream-track $\proton\pim$ categories.
The first and second uncertainties are statistical and systematic, respectively. 
} 
\label{tab:results}
\centering
\begin{tabular}{l|c|l|c}
Obs.  & Value & Obs.  & Value  \\ 
\hline
$K_{1}$ & $\phantom{+}0.346 \pm 0.020 \pm 0.004$ & $K_{18}$ & $-0.108 \pm 0.058 \pm 0.008$ \\
$K_{2}$ & $\phantom{+}0.308 \pm 0.040 \pm 0.008$ & $K_{19}$ & $-0.151 \pm 0.122 \pm 0.022$ \\
$K_{3}$ & $-0.261 \pm 0.029 \pm 0.006$ & $K_{20}$ & $-0.116 \pm 0.056 \pm 0.008$ \\
$K_{4}$ & $-0.176 \pm 0.046 \pm 0.016$ & $K_{21}$ & $-0.041 \pm 0.105 \pm 0.020$ \\
$K_{5}$ & $-0.251 \pm 0.081 \pm 0.016$ & $K_{22}$ & $-0.014 \pm 0.045 \pm 0.007$ \\
$K_{6}$ & $\phantom{+}0.329 \pm 0.055 \pm 0.012$ & $K_{23}$ & $-0.024 \pm 0.077 \pm 0.012$ \\
$K_{7}$ & $-0.015 \pm 0.084 \pm 0.013$ & $K_{24}$ & $\phantom{+}0.005 \pm 0.033 \pm 0.005$ \\
$K_{8}$ & $-0.099 \pm 0.037 \pm 0.012$ & $K_{25}$ & $-0.226 \pm 0.176 \pm 0.030$ \\
$K_{9}$ & $\phantom{+}0.005 \pm 0.084 \pm 0.012$ & $K_{26}$ & $\phantom{+}0.140 \pm 0.074 \pm 0.014$ \\
$K_{10}$ & $-0.045 \pm 0.037 \pm 0.006$ & $K_{27}$ & $\phantom{+}0.016 \pm 0.140 \pm 0.025$ \\
$K_{11}$ & $-0.007 \pm 0.043 \pm 0.009$ & $K_{28}$ & $\phantom{+}0.032 \pm 0.058 \pm 0.009$ \\
$K_{12}$ & $-0.009 \pm 0.063 \pm 0.014$ & $K_{29}$ & $-0.127 \pm 0.097 \pm 0.016$ \\
$K_{13}$ & $\phantom{+}0.024 \pm 0.045 \pm 0.010$ & $K_{30}$ & $\phantom{+}0.011 \pm 0.061 \pm 0.011$ \\
$K_{14}$ & $\phantom{+}0.010 \pm 0.082 \pm 0.013$ & $K_{31}$ & $\phantom{+}0.180 \pm 0.094 \pm 0.015$ \\
$K_{15}$ & $\phantom{+}0.158 \pm 0.117 \pm 0.027$ & $K_{32}$ & $-0.009 \pm 0.055 \pm 0.008$ \\
$K_{16}$ & $\phantom{+}0.050 \pm 0.084 \pm 0.023$ & $K_{33}$ & $\phantom{+}0.022 \pm 0.060 \pm 0.009$ \\
$K_{17}$ & $-0.000 \pm 0.120 \pm 0.022$ & $K_{34}$ & $\phantom{+}0.060 \pm 0.058 \pm 0.009$ \\
\end{tabular}
\end{table}

\begin{figure}[!tb]
\centering
\includegraphics[width=0.49\linewidth]{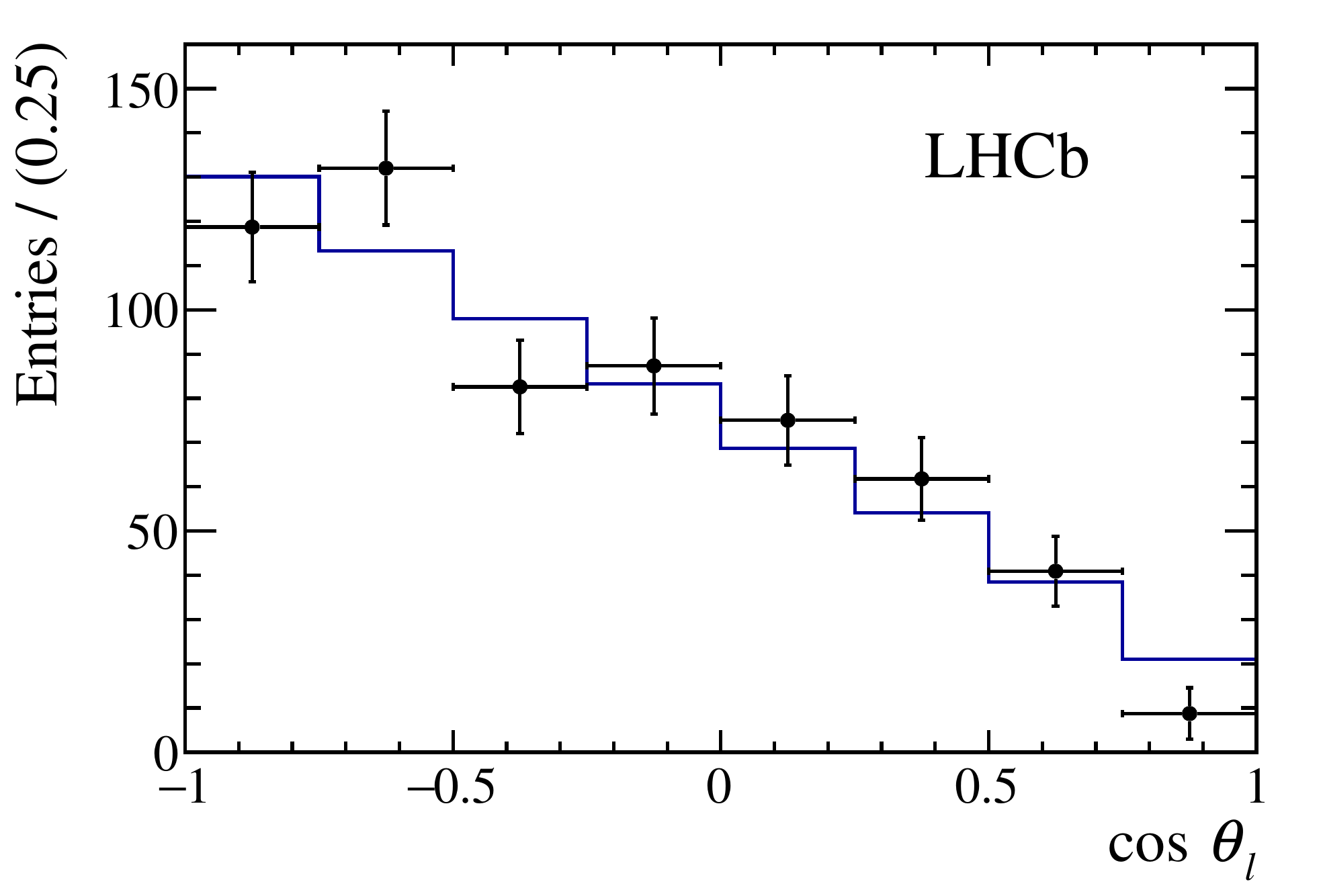} 
\includegraphics[width=0.49\linewidth]{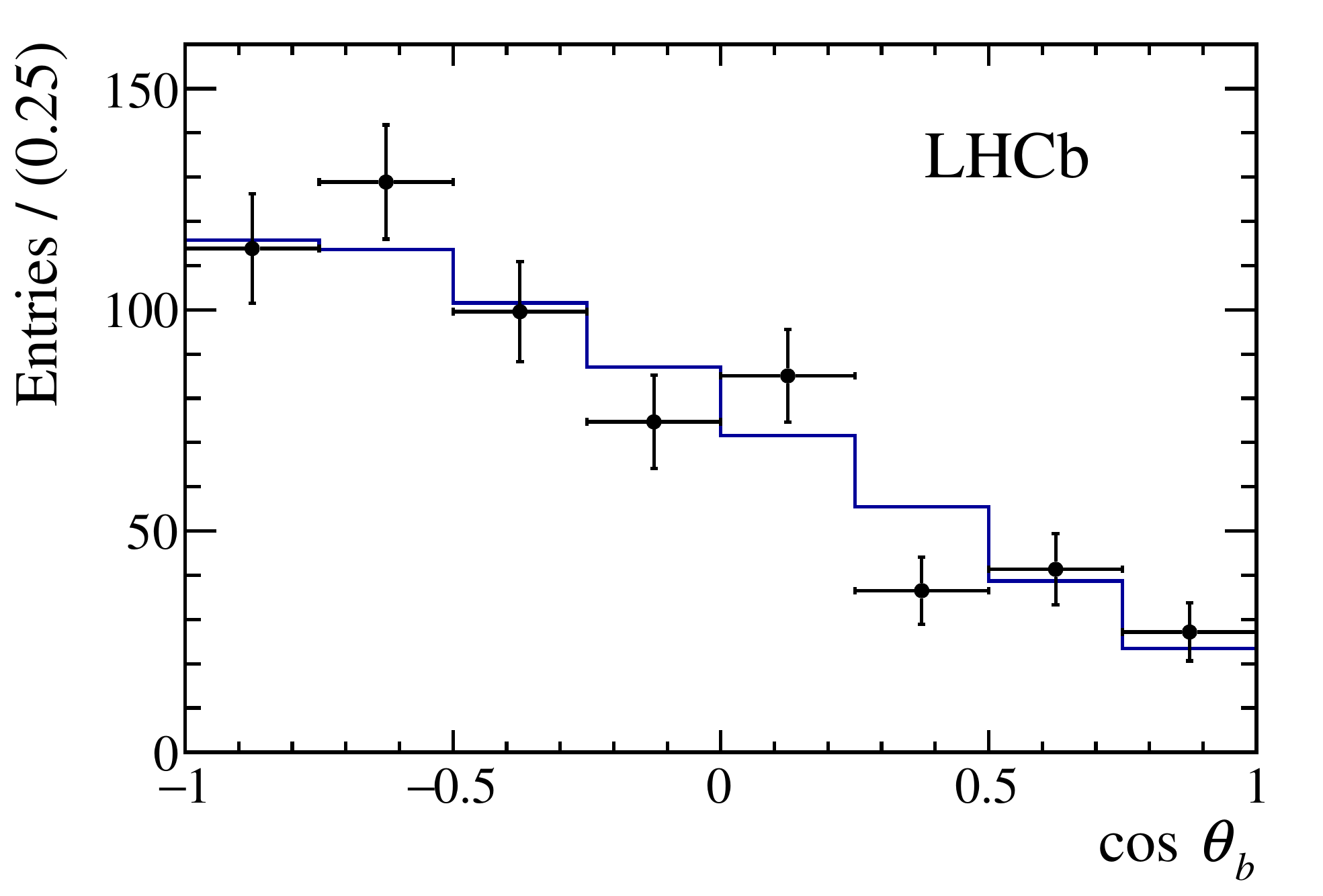} \\
\includegraphics[width=0.49\linewidth]{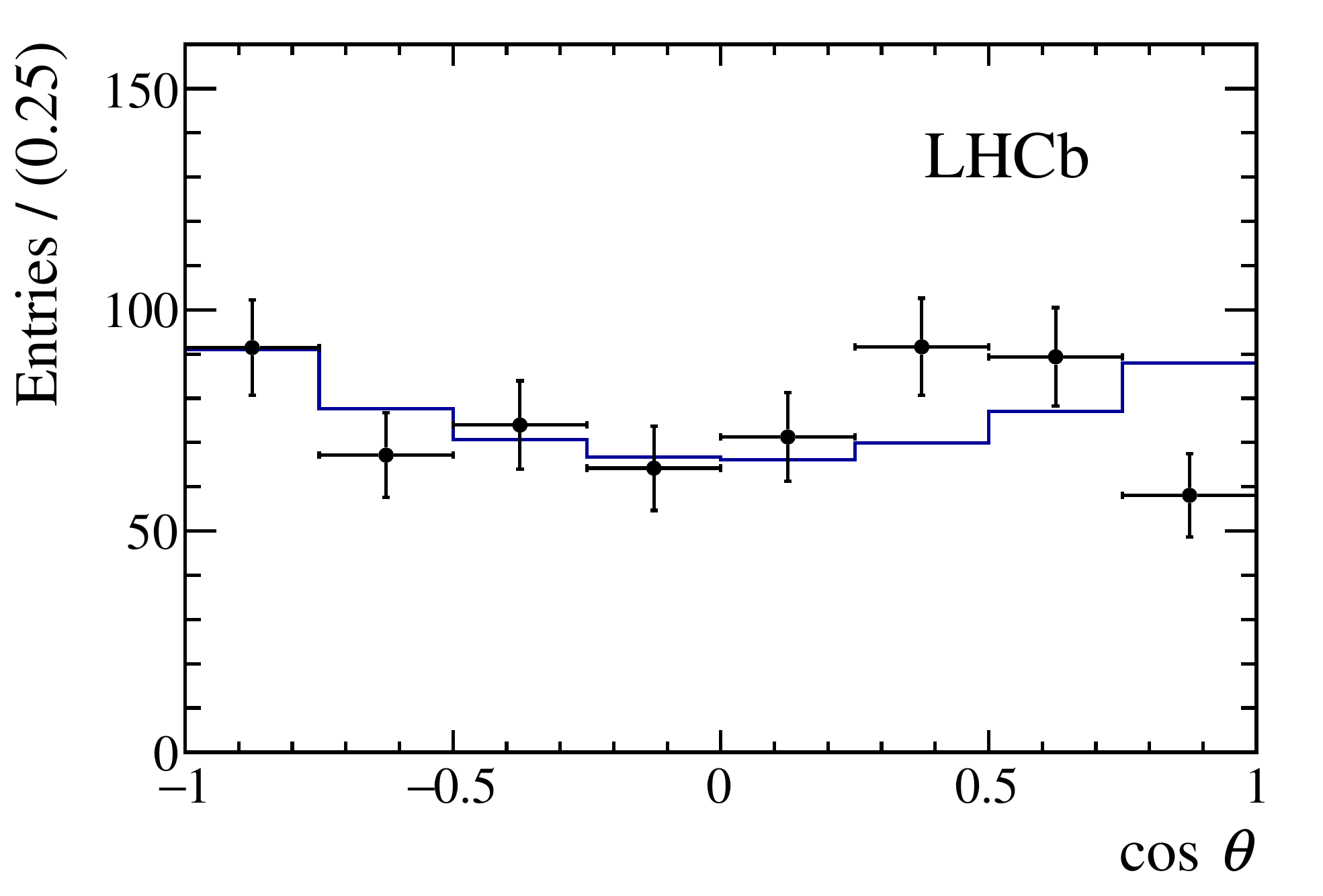} 
\includegraphics[width=0.49\linewidth]{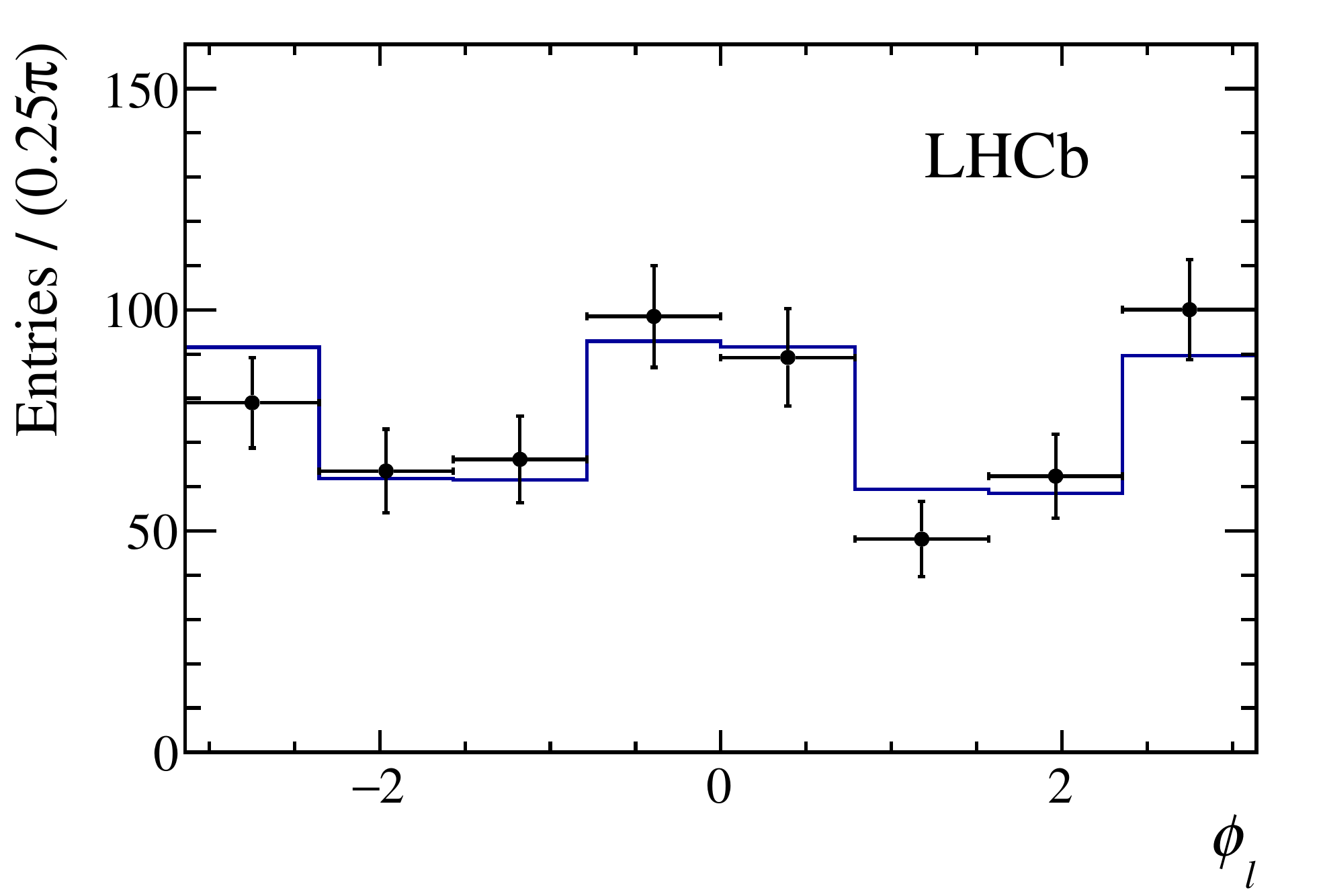} \\ 
\includegraphics[width=0.49\linewidth]{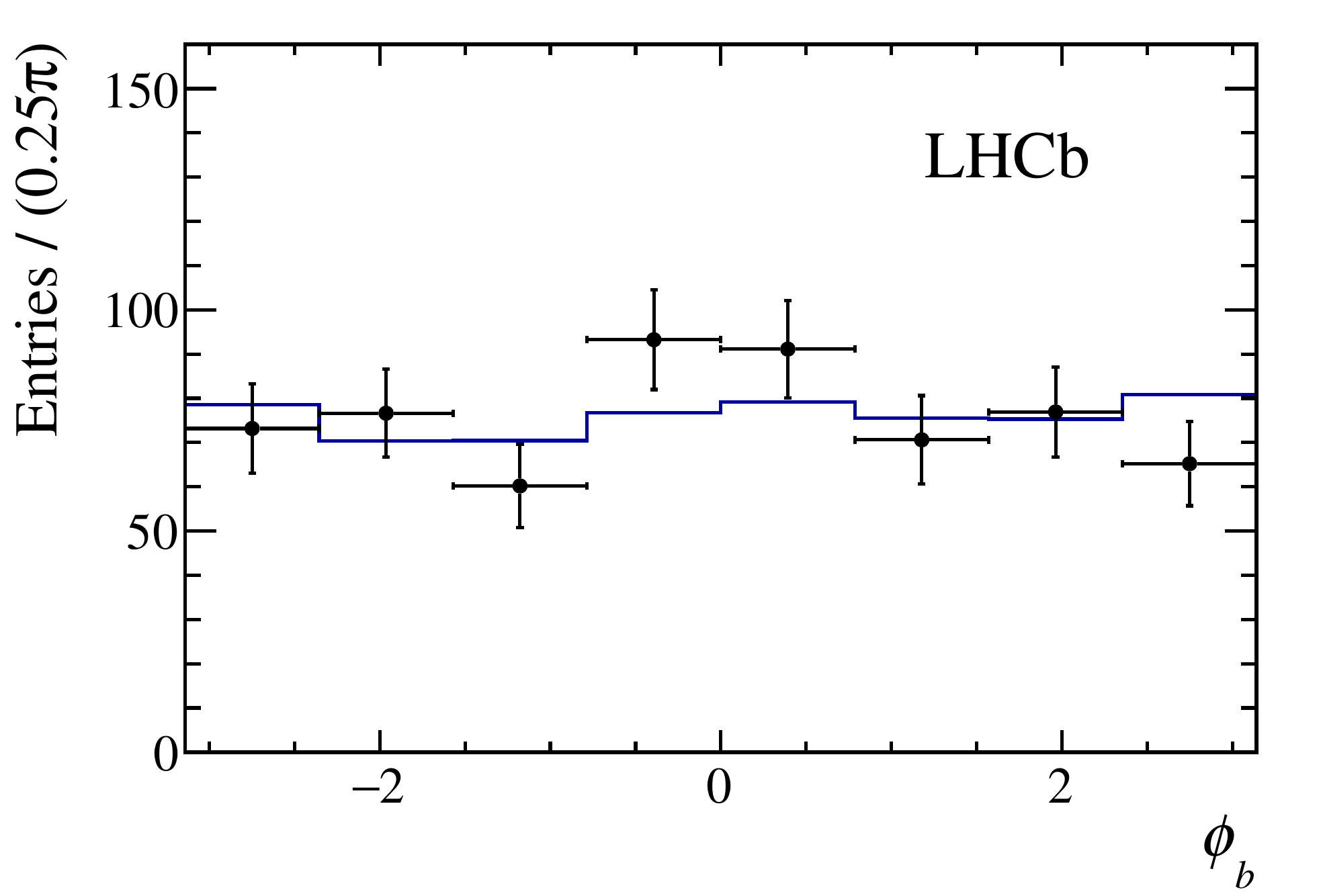} 
 \caption{ One-dimensional projections of the angular distributions of
   the candidates (black points), combining Run\,1 and Run\,2 data, as well
   as candidates reconstructed in the long- and downstream-track $p\pim$
   categories.  The background is subtracted from the data but no
   efficiency correction is applied.  The projection of
   each angular distribution obtained from the moment analysis
   multiplied by the efficiency distribution is superimposed.  
   The large variation in $\phi_\ell$ is primarily due to the angular acceptance.}
\label{fig:projections}
\end{figure}

\section{Systematic uncertainties}
\label{sec:Systematics}

The angular observables may be sensitive to systematic effects arising
from imperfect modelling of either the angular efficiency or
the $m(p\pim\mumu)$ distribution.  Where possible, systematic
uncertainties are estimated using pseudoexperiments. 
These are generated from a systematically varied model and
the observables are then estimated using the nominal analysis,
neglecting the variation in the generation.  The sources of systematic
uncertainty considered are listed in
Table~\ref{tab:systematics}.  In general, systematic
uncertainties are found to be small compared to the statistical
uncertainties on the measurements.

The largest systematic uncertainties in modelling the angular
efficiency are from the size of the simulated data samples and
the order of the Legendre polynomials used to parameterise the efficiency.  The
former is determined by
bootstrapping the simulated sample and re-evaluating the model.
The latter is estimated by increasing the order of polynomials used in the efficiency parameterisation by up to two orders.
 By default, the efficiency model is chosen to have the minimum number of terms needed to get a good description of both the simulated and the data control samples. 
 Increasing further the number of terms results in an overfitting of statistical fluctuations in the simulated data used to determine the efficiency model 
 (due to the limited size of the simulated data set).

A systematic uncertainty due to the modelling of the data by the simulation is
estimated by varying the tracking and muon identification efficiencies, and by
applying an additional correction to the \pt and $\eta$ spectra of
the \Lb baryons.

The impact of neglecting angular resolution when determining the
angular observables is estimated by smearing pseudoexperiments
according to the resolution determined using simulated data.  The angular resolution is
poorest for $\theta$, \thetab and \phib in the
downstream $p\pim$ category, at around 90\mrad for $\theta$ and \thetab and 150\mrad for \phib.

In the calculation of
the angular basis, the crossing angle of the LHC beams is neglected.  The impact of this is 
 estimated by generating pseudoexperiments with the correct crossing
 angle and neglecting this when the angular observables are determined.

The systematic uncertainty due to modelling the shape of the signal mass
distribution is small.
The main contribution to this uncertainty comes from the modelling of the 
tails of the signal mass distribution.
The factorisation of the mass model and the angular
distribution, which is a requirement of the \sPlot technique, is also tested and results in a negligible
systematic uncertainty.

\begin{table}[!tb]
\caption{Sources of systematic uncertainty on the $K_i$ angular
  observables, together with the mean and the range 
  of uncertainty values assigned to the 34 $K_i$ parameters in each case.  The variation of
  each source of systematic uncertainty between the different observables depends on the 
  structure of the weighting functions used to extract the observable and its correlation with the angular efficiency. 
  }
\label{tab:systematics}
\centering
\begin{tabular}{l|c|c}
Source & \multicolumn{2}{c}{Uncertainty [$10^{-3}$]} \\
           & Range among $K_i$  &  Mean \\ 
\hline
Simulated sample size & 3--22 & 9  \\ 
Efficiency parameterisation & 1--13 & 4   \\ 
Data-simulation differences & 2--16 & 6   \\
Angular resolution & 1--11 &  4 \\  
Beam crossing angle & 1--8 & 4 \\
Signal mass model & 1--4 & 2 \\ 
\end{tabular}
\end{table}

\section{Summary}
\label{sec:Summary}

%Summary of results

An analysis of the angular distribution of the decay
\decay{\Lb}{\Lz\mumu} in the dimuon invariant mass squared range $15 <
\qsq < 20\gev^2/c^4$ is reported.
Using data collected with the LHCb
detector between 2011 and 2016,
 the full basis of angular observables is measured for the first
 time.  From the measured observables, the
lepton-side, hadron-side and combined forward-backward asymmetries of
the decay are determined to be
\begin{displaymath}
\begin{split}
\AFBl  & = -0.39 \pm 0.04\stat \,\pm 0.01\syst~, \\ 
\AFBh  & = -0.30 \pm 0.05\stat \,\pm 0.02\syst~, \\ 
\AFBlh & = +0.25 \pm 0.04\stat \,\pm 0.01\syst~. 
\end{split}
\end{displaymath} 
The results presented here supersede the results for angular
observables in Ref.\ \cite{LHCb-PAPER-2015-009} (see discussion in Sec.~\ref{sec:Results}).  The measured
angular observables are compatible with the SM predictions obtained
using the \EOS software~\cite{OurEOSVersion}, where the
\Lb production polarisation is set to the value obtained by the LHCb
collaboration in $pp$ collisions at a centre-of-mass energy of
$7$\,TeV~\cite{LHCb-PAPER-2012-057}.

\section*{Acknowledgements}
%
% These Acknowledgements valid from 6-July-2018
%
\noindent We express our gratitude to our colleagues in the CERN
accelerator departments for the excellent performance of the LHC. We
thank the technical and administrative staff at the LHCb
institutes.
We acknowledge support from CERN and from the national agencies:
CAPES, CNPq, FAPERJ and FINEP (Brazil); 
MOST and NSFC (China); 
CNRS/IN2P3 (France); 
BMBF, DFG and MPG (Germany); 
INFN (Italy); 
NWO (Netherlands); 
MNiSW and NCN (Poland); 
MEN/IFA (Romania); 
MinES and FASO (Russia); 
MinECo (Spain); 
SNSF and SER (Switzerland); 
NASU (Ukraine); 
STFC (United Kingdom); 
NSF (USA).
We acknowledge the computing resources that are provided by CERN, IN2P3
(France), KIT and DESY (Germany), INFN (Italy), SURF (Netherlands),
PIC (Spain), GridPP (United Kingdom), RRCKI and Yandex
LLC (Russia), CSCS (Switzerland), IFIN-HH (Romania), CBPF (Brazil),
PL-GRID (Poland) and OSC (USA).
We are indebted to the communities behind the multiple open-source
software packages on which we depend.
Individual groups or members have received support from
AvH Foundation (Germany);
EPLANET, Marie Sk\l{}odowska-Curie Actions and ERC (European Union);
ANR, Labex P2IO and OCEVU, and R\'{e}gion Auvergne-Rh\^{o}ne-Alpes (France);
Key Research Program of Frontier Sciences of CAS, CAS PIFI, and the Thousand Talents Program (China);
RFBR, RSF and Yandex LLC (Russia);
GVA, XuntaGal and GENCAT (Spain);
the Royal Society
and the Leverhulme Trust (United Kingdom);
Laboratory Directed Research and Development program of LANL (USA).

\clearpage
\appendix
{\noindent\normalfont\bfseries\Large Appendices}

\section{Angular basis}
\label{appendix:Basis}

The angular distribution of the decay \decay{\Lb}{\Lz\mumu} is described by five angles, $\theta$, \thetal, \phil, \thetab and \phib defined with respect to the normal-vector 
\begin{align}
\hat{n} = \frac{\vec{p}_{\rm beam}^{\,\{{\rm lab}\}}\times \vec{p}_{\Lb}^{\,\{{\rm lab}\}}}{|\vec{p}_{\rm beam}^{\,\{{\rm lab}\}} \times \vec{p}_{\Lb}^{\,\{{\rm lab}\}}|}~,
\end{align}
where $\hat{p} = \vec{p}/|\vec{p}|$ and the parentheses refer to the rest frame the momentum is measured in.
The angle $\theta$ is defined by the angle between $\hat{n}$ and the \Lz baryon momentum in the \Lb baryon rest frame, \ie
\begin{align}
\cos\theta = \hat{n} \cdot \hat{p}_{\Lz}^{\{\Lb\}}~.
\end{align}
The decay of the \Lz baryon and the dimuon system can be described by coordinate systems with $\hat{z}_{b} = \hat{p}_{\Lz}^{\{\Lb\}}$ and $\hat{z}_{\ell} = \hat{p}_{\mumu}^{\{\Lb\}}$, $\hat{y}_{b,\ell} = \hat{n} \times \hat{z}_{b,\ell}$ and $\hat{x}_{b,\ell} = \hat{z}_{b,\ell} \times \hat{y}_{b,\ell}$. 
The angles \thetab and \phib (\thetal and \phil) are the polar and azimuthal angle of the proton (\mup) in the \Lz baryon (dimuon) rest frame. 
The angles are defined by
 \begin{align}
 \begin{split}
\cos\theta_{b,\ell} &= \hat{z}_{b,\ell}\cdot\hat{p}_{b,\ell}  ~,~\cos\phi_{b,\ell}  = \hat{x}_{b,\ell} \cdot \hat{p}_{\perp b,\ell}~,~\sin\phi_{b,\ell} = \hat{y}_{b,\ell} \cdot \hat{p}_{\perp b,\ell}~,
\end{split}
\end{align}
where $\hat{p}_{b}$ $(\hat{p}_{\ell})$ is the direction of the proton
(\mup) and $\hat{p}_{\perp b}$ $(\hat{p}_{\perp\ell})$ is a unit
vector corresponding to the component perpendicular to the $\hat{z}_b$
$(\hat{z}_{\ell})$ axis.
For the \Lbbar decay, the angular variables are transformed such that $\thetal  \to \pi-\thetal$, $\phil \to \pi - \phil$ and $\phib \to -\phib$.
This ensures that, in the absence of \CP violating effects, the $K_{i}$ observables are the same for \Lb and \Lbbar decays. 

\section{Angular distribution and weighting functions}
\label{appendix:distribution}

The full form of the angular distribution of the decay \decay{\Lb}{\Lz\mumu} is given by
\begin{align}
%\begin{split}
\frac{\deriv^{5}\Gamma}{\deriv\vec{\Omega}} = \frac{3}{32\pi^{2}} \Big(
& \left(K_1\sin^2\theta_\ell+K_2\cos^2\theta_\ell+K_3\cos\theta_\ell\right)  +  \\[-5pt]
& \left(K_4\sin^2\theta_\ell+K_5\cos^2\theta_\ell+K_6\cos\theta_\ell\right)\cos\theta_b + \nonumber \\
& \left(K_7\sin\theta_\ell\cos\theta_\ell+K_8\sin\theta_\ell\right)\sin\theta_b\cos\left(\phi_b+\phi_\ell\right) + \nonumber \\
&\left(K_9\sin\theta_\ell\cos\theta_\ell+K_{10}\sin\theta_\ell\right)\sin\theta_b\sin\left(\phi_b+\phi_\ell\right) + \nonumber \\
& \left(K_{11}\sin^2\theta_\ell+K_{12}\cos^2\theta_\ell+K_{13}\cos\theta_\ell\right)\cos\theta + \nonumber \\
& \left( K_{14}\sin^2\theta_\ell+K_{15}\cos^2\theta_\ell+K_{16}\cos\theta_\ell\right)\cos\theta_b \cos\theta + \nonumber  \\
& \left(K_{17}\sin\theta_\ell\cos\theta_\ell+K_{18}\sin\theta_\ell\right)\sin\theta_b\cos\left(\phi_b+\phi_\ell\right)\cos\theta  + \nonumber \\
& \left(K_{19}\sin\theta_\ell\cos\theta_\ell+K_{20}\sin\theta_\ell\right)\sin\theta_b\sin\left(\phi_b+\phi_\ell\right) \cos\theta + \nonumber \\
& \left(K_{21}\cos\theta_\ell\sin\theta_\ell+K_{22}\sin\theta_\ell\right)\sin\phi_\ell \sin\theta + \nonumber \\
& \left(K_{23}\cos\theta_\ell\sin\theta_\ell+K_{24}\sin\theta_\ell\right)\cos\phi_\ell  \sin\theta + \nonumber \\
& \left(K_{25}\cos\theta_\ell\sin\theta_\ell+K_{26}\sin\theta_\ell\right)\sin\phi_\ell\cos\theta_b  \sin\theta + \nonumber \\
& \left(K_{27}\cos\theta_\ell\sin\theta_\ell+K_{28}\sin\theta_\ell\right)\cos\phi_\ell\cos\theta_b  \sin\theta  + \nonumber \\
& \left(K_{29}\cos^2\theta_\ell+K_{30}\sin^2\theta_\ell\right)\sin\theta_b\sin\phi_b  \sin\theta + \nonumber \\
& \left(K_{31}\cos^2\theta_\ell+K_{32}\sin^2\theta_\ell\right)\sin\theta_b\cos\phi_b  \sin\theta + \nonumber \\
& \left(K_{33}\sin^2\theta_\ell \right) \sin\theta_b\cos\left(2\phi_\ell+\phi_b\right) \sin\theta  + \nonumber \\
& \left(K_{34}\sin^2\theta_\ell \right) \sin\theta_b\sin\left(2\phi_\ell+\phi_b\right)  \sin\theta  \Big)~.\nonumber
%\end{split}
\label{eq:appendix:34terms}
\end{align} 
The individual $K_{i}$ parameters can be determined using a moment analysis of the angular distribution with the weighting functions 
\begin{align}
g_{1}(\vec{\Omega}) = & \tfrac{1}{4}(3-5\cos^2\theta_\ell)~,  
&&  g_{6}(\vec{\Omega}) ~= 3\cos\theta_\ell\cos\theta_b ~,    \\ 
g_{2}(\vec{\Omega}) = & \tfrac{1}{2}(5\cos^2\theta_\ell-1)~,  
&& g_{7}(\vec{\Omega}) ~= \tfrac{15}{2}\cos\theta_\ell\sin\theta_\ell\sin\theta_b\cos(\phi_\ell + \phi_b)~,   \nonumber  \\
g_{3}(\vec{\Omega}) = & \cos\theta_\ell ~,  
&& g_{8}(\vec{\Omega}) ~= \tfrac{3}{2}\sin\theta_\ell\sin\theta_b\cos(\phi_\ell + \phi_b)~, \nonumber \\ 
g_{4}(\vec{\Omega}) = & \tfrac{3}{4}(3 - 5 \cos^{2}\theta_\ell) \cos\theta_b ~, 
&& g_{9}(\vec{\Omega}) ~= \tfrac{15}{2} \cos\theta_\ell\sin\theta_\ell\sin\theta_b\sin(\phi_\ell + \phi_b)~,  \nonumber  \\
g_{5}(\vec{\Omega}) = & \tfrac{3}{2}( 5\cos^2\theta_\ell - 1) \cos\theta_b~,
&& g_{10}(\vec{\Omega}) = \tfrac{3}{2} \sin\theta_\ell\sin\theta_b\sin(\phi_\ell + \phi_b)~,  \nonumber
\end{align}
from Ref.~\cite{Beaujean:2015xea} and the weighting functions 
\begin{align}
g_{11}(\vec{\Omega}) =  & \tfrac{3}{4}(3-5\cos^2\theta_\ell)\cos\theta ~,
&& g_{23}(\vec{\Omega}) =  \tfrac{15}{2} \cos\theta_\ell\sin\theta_\ell\sin\theta\cos\phi_\ell~,   \\ 
g_{12}(\vec{\Omega}) =  & \tfrac{3}{2}(5\cos^2\theta_\ell-1)\cos\theta ~,
&& g_{24}(\vec{\Omega}) = \tfrac{3}{2} \sin\theta\sin\theta_\ell\cos\phi_\ell ~, \nonumber \\
g_{13}(\vec{\Omega}) =  & 3\cos\theta_\ell\cos\theta~, 
&& g_{25}(\vec{\Omega}) =  \tfrac{45}{2}\cos\theta_\ell\sin\theta_\ell\cos\theta_b\sin\theta\sin\phi_\ell ~, \nonumber \\ 
g_{14}(\vec{\Omega}) =  & \tfrac{9}{4}(3 - 5\cos^{2}\theta_\ell) \cos\theta_b \cos\theta~, 
&& g_{26}(\vec{\Omega}) = \tfrac{9}{2}\sin\theta\sin\theta_\ell\cos\theta_b\sin\phi_\ell~, \nonumber \\
g_{15}(\vec{\Omega}) =  & \tfrac{9}{2} ( 5\cos^2\theta_\ell - 1 ) \cos\theta_b\cos\theta~,
&& g_{27}(\vec{\Omega}) =  \tfrac{45}{2}\cos\theta_\ell\sin\theta_\ell\cos\theta_b\sin\theta\cos\phi_\ell~, \nonumber \\
g_{16}(\vec{\Omega}) =  & 9\cos\theta\cos\theta_\ell\cos\theta_b ~,
&& g_{28}(\vec{\Omega}) = \tfrac{9}{2} \sin\theta\sin\theta_\ell\cos\theta_b\cos\phi_\ell ~, \nonumber \\
g_{17}(\vec{\Omega}) =  & \tfrac{45}{2} \cos\theta_\ell\sin\theta_\ell\sin\theta_b\cos\theta\cos(\phi_\ell + \phi_b)~, 
&& g_{29}(\vec{\Omega}) = \tfrac{9}{4}(5\cos^2\theta_\ell - 1)\sin\theta_b\sin\theta\sin\phi_b~, \nonumber \\
g_{18}(\vec{\Omega}) =  & \tfrac{9}{2}\sin\theta_\ell\sin\theta_b\cos\theta\cos(\phi_\ell + \phi_b)~, 
&& g_{30}(\vec{\Omega}) =  \tfrac{9}{8}(3 - 5\cos^2\theta_\ell)\sin\theta_b\sin\theta\sin\phi_b~,  \nonumber \\
g_{19}(\vec{\Omega}) =  & \tfrac{45}{2}\cos\theta_\ell\sin\theta_\ell\sin\theta_b\cos\theta\sin(\phi_\ell+\phi_b)~, 
&&  g_{31}(\vec{\Omega}) =  \tfrac{9}{4}(5 \cos^2\theta_\ell - 1)\sin\theta_b\sin\theta\cos\phi_b~, \nonumber \\
g_{20}(\vec{\Omega}) =  & \tfrac{9}{2}\sin\theta_\ell\sin\theta_b\cos\theta\sin(\phi_\ell + \phi_b)~, 
&& g_{32}(\vec{\Omega}) =  \tfrac{9}{8}(3 - 5\cos^2\theta_\ell)\sin\theta_b\sin\theta\cos\phi_b~, \nonumber \\
g_{21}(\vec{\Omega}) =  & \tfrac{15}{2}\cos\theta_\ell\sin\theta_\ell\sin\theta\sin\phi_\ell ~,
&& g_{33}(\vec{\Omega}) =  \tfrac{9}{4}\sin\theta_b\sin\theta\cos(2\phi_\ell + \phi_b)~, \nonumber \\
g_{22}(\vec{\Omega}) =  & \tfrac{3}{2} \sin\theta\sin\theta_\ell\sin\phi_\ell~,  
&& g_{34}(\vec{\Omega}) =  \tfrac{9}{4}\sin\theta_b\sin\theta\sin(2\phi_\ell + \phi_b)~, \nonumber 
\end{align}
from Ref.~\cite{Blake:2017une}.

\section{Results separated by data-taking period}
\label{appendix:results:split}

Tables~\ref{tab:results:run1} and \ref{tab:results:run2} show the
values of the observables for each of the two data-taking periods.
Table~\ref{tab:results:run1} shows the values of the observables
combining the 2011 data, collected at $\sqs=7\tev$, and the
2012 data, collected at $\sqs=8\tev$.
Table~\ref{tab:results:run2} shows the values of the observables in
the Run\,2 data, collected at $\sqs=13\tev$.  

\begin{table}[!htb]
\protect\caption{Measured values for the angular observables from the Run\,1 data 
combining the results of the moments obtained from the candidates reconstructed in
the long- and downstream-track $\proton\pim$ categories.
The first and second uncertainties are statistical and systematic, respectively. 
}
\centering
\begin{tabular}{l|c|l|c}
Obs.  & Value & Obs.  & Value  \\ 
\hline
$K_{1}$ & $\phantom{+}0.376 \pm 0.029 \pm 0.006$ & $K_{18}$ & $-0.081 \pm 0.081 \pm 0.015$ \\
$K_{2}$ & $\phantom{+}0.248 \pm 0.057 \pm 0.012$ & $K_{19}$ & $-0.023 \pm 0.165 \pm 0.031$ \\
$K_{3}$ & $-0.241 \pm 0.041 \pm 0.008$ & $K_{20}$ & $-0.156 \pm 0.078 \pm 0.019$ \\
$K_{4}$ & $-0.212 \pm 0.070 \pm 0.013$ & $K_{21}$ & $-0.050 \pm 0.150 \pm 0.032$ \\
$K_{5}$ & $-0.123 \pm 0.117 \pm 0.020$ & $K_{22}$ & $\phantom{+}0.032 \pm 0.064 \pm 0.014$ \\
$K_{6}$ & $\phantom{+}0.247 \pm 0.079 \pm 0.017$ & $K_{23}$ & $\phantom{+}0.038 \pm 0.104 \pm 0.018$ \\
$K_{7}$ & $-0.027 \pm 0.124 \pm 0.022$ & $K_{24}$ & $\phantom{+}0.004 \pm 0.047 \pm 0.008$ \\
$K_{8}$ & $-0.081 \pm 0.054 \pm 0.010$ & $K_{25}$ & $-0.107 \pm 0.254 \pm 0.046$ \\
$K_{9}$ & $-0.123 \pm 0.115 \pm 0.018$ & $K_{26}$ & $\phantom{+}0.130 \pm 0.106 \pm 0.024$ \\
$K_{10}$ & $\phantom{+}0.021 \pm 0.051 \pm 0.009$ & $K_{27}$ & $-0.200 \pm 0.190 \pm 0.035$ \\
$K_{11}$ & $-0.030 \pm 0.062 \pm 0.014$ & $K_{28}$ & $\phantom{+}0.058 \pm 0.084 \pm 0.015$ \\
$K_{12}$ & $-0.114 \pm 0.092 \pm 0.022$ & $K_{29}$ & $-0.172 \pm 0.142 \pm 0.027$ \\
$K_{13}$ & $\phantom{+}0.059 \pm 0.064 \pm 0.016$ & $K_{30}$ & $-0.060 \pm 0.088 \pm 0.014$ \\
$K_{14}$ & $\phantom{+}0.122 \pm 0.126 \pm 0.026$ & $K_{31}$ & $\phantom{+}0.252 \pm 0.126 \pm 0.022$ \\
$K_{15}$ & $\phantom{+}0.247 \pm 0.171 \pm 0.042$ & $K_{32}$ & $-0.074 \pm 0.075 \pm 0.011$ \\
$K_{16}$ & $-0.193 \pm 0.116 \pm 0.029$ & $K_{33}$ & $-0.010 \pm 0.081 \pm 0.014$ \\
$K_{17}$ & $-0.119 \pm 0.178 \pm 0.033$ & $K_{34}$ & $\phantom{+}0.140 \pm 0.088 \pm 0.012$ \\
\end{tabular}
\label{tab:results:run1}
\end{table}

\begin{table}[!htb]
\protect\caption{Measured values for the angular observables from the Run\,2 data 
combining the results of the moments obtained from the candidates reconstructed in
the long- and downstream-track $\proton\pim$ categories.
The first and second uncertainties are statistical and systematic, respectively. 
}
\centering
\begin{tabular}{l|c|l|c}
Obs.  &  Value & Obs.  &  Value  \\
\hline
$K_{1}$ & $\phantom{+}0.318 \pm 0.028 \pm 0.007$ & $K_{18}$ & $-0.134 \pm 0.081 \pm 0.014$ \\
$K_{2}$ & $\phantom{+}0.364 \pm 0.056 \pm 0.013$ & $K_{19}$ & $-0.273 \pm 0.178 \pm 0.040$ \\
$K_{3}$ & $-0.279 \pm 0.042 \pm 0.010$ & $K_{20}$ & $-0.078 \pm 0.082 \pm 0.017$ \\
$K_{4}$ & $-0.143 \pm 0.063 \pm 0.012$ & $K_{21}$ & $-0.033 \pm 0.142 \pm 0.023$ \\
$K_{5}$ & $-0.372 \pm 0.113 \pm 0.024$ & $K_{22}$ & $-0.058 \pm 0.062 \pm 0.008$ \\
$K_{6}$ & $\phantom{+}0.407 \pm 0.076 \pm 0.017$ & $K_{23}$ & $-0.082 \pm 0.111 \pm 0.018$ \\
$K_{7}$ & $-0.004 \pm 0.114 \pm 0.018$ & $K_{24}$ & $\phantom{+}0.005 \pm 0.046 \pm 0.008$ \\
$K_{8}$ & $-0.116 \pm 0.051 \pm 0.011$ & $K_{25}$ & $-0.339 \pm 0.243 \pm 0.042$ \\
$K_{9}$ & $\phantom{+}0.126 \pm 0.124 \pm 0.017$ & $K_{26}$ & $\phantom{+}0.150 \pm 0.101 \pm 0.017$ \\
$K_{10}$ & $-0.108 \pm 0.054 \pm 0.008$ & $K_{27}$ & $\phantom{+}0.221 \pm 0.203 \pm 0.036$ \\
$K_{11}$ & $\phantom{+}0.014 \pm 0.060 \pm 0.009$ & $K_{28}$ & $\phantom{+}0.008 \pm 0.083 \pm 0.015$ \\
$K_{12}$ & $\phantom{+}0.091 \pm 0.085 \pm 0.023$ & $K_{29}$ & $-0.085 \pm 0.135 \pm 0.025$ \\
$K_{13}$ & $-0.009 \pm 0.063 \pm 0.016$ & $K_{30}$ & $\phantom{+}0.079 \pm 0.084 \pm 0.014$ \\
$K_{14}$ & $-0.096 \pm 0.105 \pm 0.019$ & $K_{31}$ & $\phantom{+}0.113 \pm 0.140 \pm 0.021$ \\
$K_{15}$ & $\phantom{+}0.073 \pm 0.159 \pm 0.046$ & $K_{32}$ & $\phantom{+}0.053 \pm 0.080 \pm 0.012$ \\
$K_{16}$ & $\phantom{+}0.280 \pm 0.120 \pm 0.034$ & $K_{33}$ & $\phantom{+}0.052 \pm 0.088 \pm 0.011$ \\
$K_{17}$ & $\phantom{+}0.112 \pm 0.166 \pm 0.033$ & $K_{34}$ & $-0.015 \pm 0.079 \pm 0.012$ \\
\end{tabular}
\label{tab:results:run2}
\end{table}

\clearpage

\section{Correlation matrices} 
\label{appendix:correlation} 

\begin{figure}[!hbt]
\centering
\includegraphics[width=0.7\linewidth]{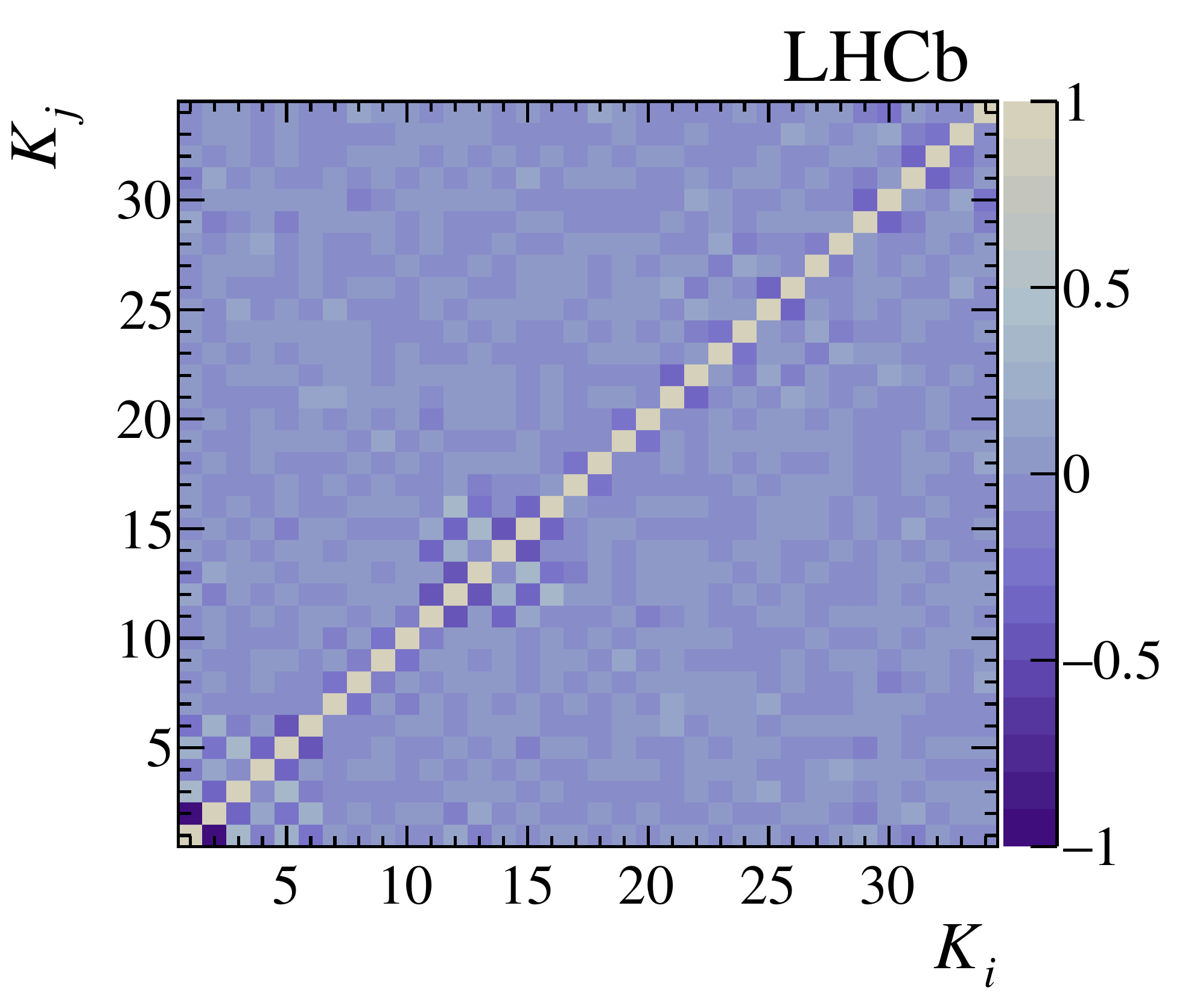}
\caption{Correlation coefficients between observables
  for the combined Run\,1 and Run\,2 data.  The correlations are
  estimated using bootstrapped samples.    
  }
\label{fig:corr:combined}
\end{figure}

\begin{figure}[!hbt]
\centering
\includegraphics[width=0.7\linewidth]{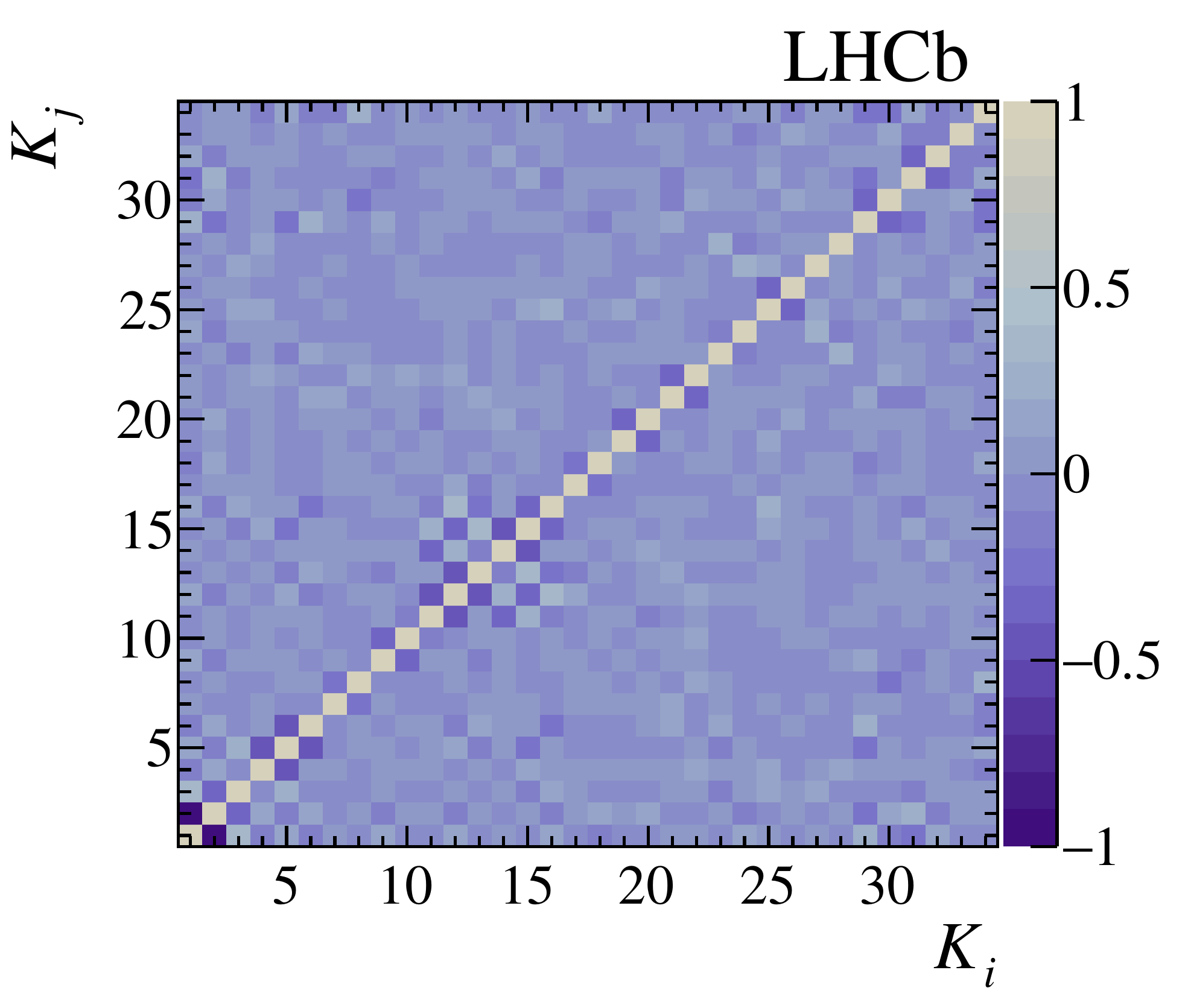}
\includegraphics[width=0.7\linewidth]{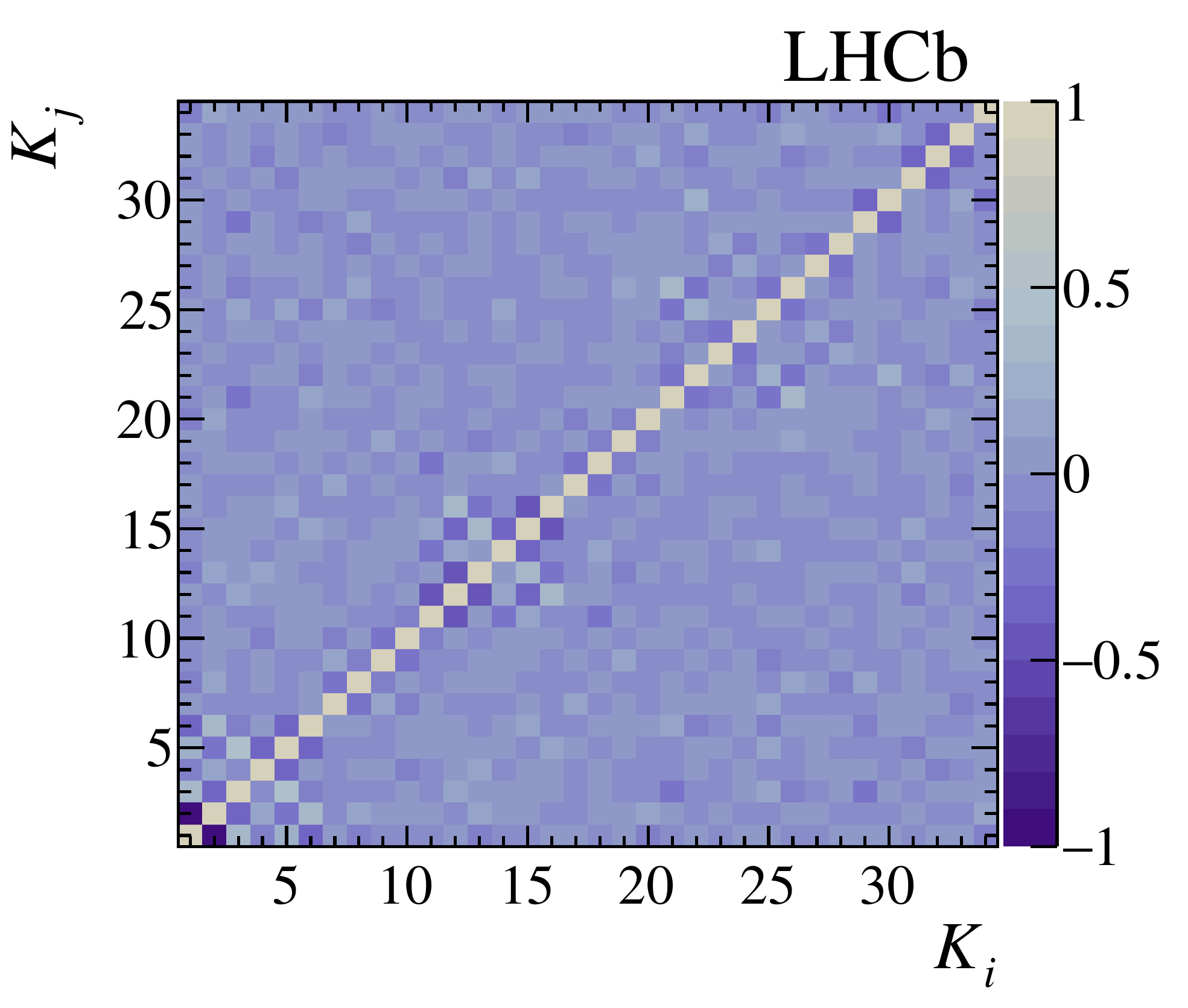}
\caption{Correlation coefficients between the observables
  (top) in the Run\,1 data and (bottom) in the Run\,2 data.
  The correlations are estimated using bootstrapped samples.
}
\label{fig:corr:runs}
\end{figure}

Figures~\ref{fig:corr:combined} and \ref{fig:corr:runs} shows the statistical correlation
between the angular observables determined using 
bootstrapped samples.  The correlation coefficients are typically small
but can be as large as 30--40\% between pairs of observables. 
The observables $K_1$ and $K_2$ are fully anticorrelated due to the normalisation of
  the observables, which requires $2 K_1 + K_2 = 1$.
The correlation matrices in numerical form are attached as supplementary material to this article.

\clearpage

\addcontentsline{toc}{section}{References}
\setboolean{inbibliography}{true}
\bibliographystyle{LHCb}
\bibliography{main,LHCb-PAPER,LHCb-CONF,LHCb-DP,LHCb-TDR}

\newpage

% Author List ----------------------------                                                                                                                                                                                                                                                                                                
\newpage
\centerline{\large\bf LHCb collaboration}
\begin{flushleft}
\small
R.~Aaij$^{27}$,
B.~Adeva$^{41}$,
M.~Adinolfi$^{48}$,
C.A.~Aidala$^{73}$,
Z.~Ajaltouni$^{5}$,
S.~Akar$^{59}$,
P.~Albicocco$^{18}$,
J.~Albrecht$^{10}$,
F.~Alessio$^{42}$,
M.~Alexander$^{53}$
% ...

{\footnotesize \it
% ...
$ ^{5}$Clermont Universit{\'e}, Universit{\'e} Blaise Pascal, CNRS/IN2P3, LPC, Clermont-Ferrand, France\\
% ...
$ ^{10}$Fakult{\"a}t Physik, Technische Universit{\"a}t Dortmund, Dortmund, Germany\\
% ...
$ ^{18}$INFN Laboratori Nazionali di Frascati, Frascati, Italy\\
% ...
$ ^{27}$Nikhef National Institute for Subatomic Physics, Amsterdam, Netherlands\\
% ...
$ ^{41}$Instituto Galego de F{\'\i}sica de Altas Enerx{\'\i}as (IGFAE), Universidade de Santiago de Compostela, Santiago de Compostela, Spain\\
$ ^{42}$European Organization for Nuclear Research (CERN), Geneva, Switzerland\\
% ...
$ ^{48}$H.H. Wills Physics Laboratory, University of Bristol, Bristol, United Kingdom\\
% ...
$ ^{53}$School of Physics and Astronomy, University of Glasgow, Glasgow, United Kingdom\\
% ...
$ ^{59}$University of Cincinnati, Cincinnati, OH, United States\\
% ...
$ ^{61}$Syracuse University, Syracuse, NY, United States\\
% ...
$ ^{73}$University of Michigan, Ann Arbor, United States, associated to $^{61}$\\
% ...
}
\end{flushleft}
\end{document}